%% file: main.tex
\documentclass[a4paper,USenglish,cleveref,autoref]{lipics-v2021}

\usepackage{booktabs}
\hideLIPIcs
\nolinenumbers

\makeatletter
\renewcommand\paragraph{\@startsection{paragraph}{4}{\z@}
  {-1.5ex \@plus-.5ex \@minus-.2ex}
  {-1em}
  {\sffamily\normalsize\bfseries}}
\makeatother

\usepackage{enumitem}
\setlist{topsep=2pt, partopsep=0pt, parsep=1pt, itemsep=1pt}
\title{EVM Workloads in the Wild: Evidence for Multi-Dimensional Gas Metering, State Growth, Delayed Execution, and Parallelism}
\titlerunning{EVM Workloads in the Wild}

\author{Lioba Heimbach}{Category Labs, Zurich, Switzerland}{liobaheimbach.cs@gmail.com}{https://orcid.org/0000-0002-8258-1712}{}
\author{Kushal Babel}{Category Labs, New York, NY, USA}{babel@cs.cornell.edu}{https://orcid.org/0000-0002-6776-5273}{}
\author{Jason Milionis}{Category Labs, New York, NY, USA}{jm@cs.columbia.edu}{https://orcid.org/0000-0002-9460-9559}{}
\authorrunning{L. Heimbach, K. Babel, J. Milionis}
\Copyright{Lioba Heimbach, Kushal Babel, Jason Milionis}


\begin{document}

\maketitle

\begin{abstract}
Gas metering on EVM-compatible blockchains assumes that execution conditions are stable: that the resource mix is constant enough to justify collapsing execution costs into a single scalar with fixed relative prices, and that state drift between submission and execution time does not materially alter a transaction's outcome. We measure the extent to which this assumption fails.

We present a trace-level measurement study of EVM workloads on Ethereum (L1) and Base (L2) throughout 2025, sampling 3{,}000 blocks per day per chain. We decompose each transaction into opcode-level execution gas, intrinsic gas, refunds, and persistent state deltas including storage slots, contract bytecode, and account state. To measure state sensitivity, we re-execute transactions sampled during September 2025 on progressively older blockchain states and record how gas usage, execution outcomes, and storage access patterns change.

We find the resource mix to be far from stable: on Base, storage reads and compute account for 29.2\% and 24.3\% of execution gas, while Ethereum devotes 34.9\% to storage writes. The mix is not stable on the same chain either: Ethereum's gas limit doubling during 2025 shifted its resource profile measurably toward more compute-heavy, Base-like patterns. Base also exhibits a higher fraction of cold storage reads at 49.7\%, compared to 39.6\% on Ethereum. Persistent state growth, a permanent cost priced as a transient one, reaches 435~GB on Base versus 30~GB on Ethereum, with different composition.

We further find that execution outcomes are equally unstable: gas estimates vary across nearby historical states for 46.0\% of transactions on Base, compared to 13.9\% on Ethereum, with especially high sensitivity for MEV and DeFi activity. Storage access patterns also diverge across execution states, limiting the effectiveness of access lists and complicating parallel execution.

Our measurements provide an empirical foundation for multi-dimensional gas metering and explicit pricing of state growth. They show that state-sensitive execution behavior complicates workload estimation and transaction parameterization, directly affecting the predictability of transactions' execution and user experience.
\end{abstract}

\input{intro}
\input{background}
\input{related_work}
\input{analysis}
\input{suggestions}

\bibliographystyle{plainurl}
\bibliography{refs}

\appendix

\input{appendix}

\end{document}

%% file: intro.tex
\section{Introduction}
\label{sec:intro}

Executing a transaction on an EVM-compatible blockchain consumes resources of fundamentally different kinds. Some are transient, i.e., constrained within a block and capacity is reset each block, such as compute, I/O, and bandwidth. Others are permanent: persistent state, i.e., accounts, contract storage, and bytecode, accumulates indefinitely on every node. Gas metering collapses all of these into a single scalar through fixed relative prices, a design that holds only if the resource mix stays constant. A second assumption runs through the transaction submission pipeline: that a transaction behaves at execution as it did in simulation, and that state drift between the two points will not materially change what it accesses, how much gas it uses, or whether it succeeds. Both assumptions are under strain.

\paragraph*{Assumption 1 violated: The workload mix shifts.}
Ethereum meters opcode costs in a single unit of account called \emph{gas}, fixing the relative prices of opcodes (cf.\ Section~\ref{sec:background}).
EVM-compatible Layer-2 (L2) systems like Base inherit this mechanism unchanged, even though their lower fees and higher throughput drive substantially different on-chain usage patterns than Layer-1 (L1) Ethereum.\footnote{While L2 transactions also incur additional costs related to posting data to Ethereum L1, these costs are orthogonal to execution on the L2 and are outside the scope of this work.}
An appropriate \emph{fixed} mapping from resource usage to relative prices is not necessarily known in advance. More fundamentally, the resource mix itself changes both across chains and over time on the same chain.
We find this concretely on Base: its workload is read- and compute-heavy where Ethereum's is write-heavy, and Base's persistent state growth is over an order of magnitude higher than Ethereum's, exceeding what differences in gas limits alone would predict. This state is permanent, yet priced under the same fixed scalar as transient resources.
Ethereum has historically adjusted opcode prices and storage rules after discovering that certain operations were systematically mispriced, particularly storage-related opcodes such as \texttt{SLOAD} and \texttt{SSTORE}~\cite{eip150,eip1884,eip2929,eip2200,eip3529}; however, the protocol is not built to react to every workload shift.
Thus, collapsing a high-dimensional workload into scalar gas is inherently brittle.

\paragraph*{Assumption 2 violated: The state moves between simulation and execution.}
High-throughput systems also change the landscape of opcode pricing because many execution costs depend on the blockchain state. For example, the gas cost of storage writes depends on whether a storage slot already exists and on whether its value transitions between zero and non-zero, making execution costs sensitive to prior state. Control-flow paths also vary with storage and account values, creating execution-context dependencies.

In high-throughput settings, rapid state updates also increase uncertainty about the state a transaction will encounter at execution time. Since gas usage must be estimated and a gas limit specified prior to submission, differences between the estimation state and the execution state can lead to variation in gas usage or even transaction failure.

Emerging execution models further amplify the consequences of such uncertainty. Under delayed or asynchronous execution, incorrect gas limit estimation can incur direct penalties, as charges may be levied based on the declared gas limit rather than actual execution, unlike in synchronous execution where overestimation is largely harmless~\cite{categorylabs_monad_initial_spec_proposal,acp194}. For users, high throughput means many transactions may intervene between submission and execution.

As a result, the same transaction may consume different gas, touch different storage keys, or fail when evaluated on different historical states, complicating workload estimation and reducing transaction predictability, with direct consequences for user experience. State-dependent access patterns further hinder mechanisms that rely on advance knowledge of accessed state, such as access lists, and increase the cost of retries in optimistic parallel execution, where re-execution may occur on a different state.

\subsection{Our Contributions}

Our goal is to clarify current EVM workload patterns and their dependence on execution state. To this end, we conduct a measurement study on Ethereum L1 and Base, an EVM-compatible optimistic rollup with the largest transaction volume and total value locked (TVL) among L2s~\cite{coingecko_layer2}. Studying these two systems allows us to contrast execution behavior under differing throughput and block-time regimes while keeping the execution environment (the EVM and its opcode pricing) identical. We pursue the following research questions:

\begin{enumerate}
    \item Workload mix: what are the differences in opcode patterns and caching behavior, as reflected by cold versus warm access rates, between Ethereum L1 and Base L2?
    \item Persistent state growth: how quickly does persistent state grow in practice, and how is this growth split between storage slots and contract bytecode?
    \item State sensitivity: how sensitive are transaction execution results to evaluation on nearby historical states relative to the state at which they actually executed?
\end{enumerate}

\noindent To answer these questions, we make the following tactical contributions:

\begin{enumerate}
\item \emph{Trace-level decomposition.}
We sample 3{,}000 blocks per day per chain and decompose each transaction into:
(i) opcode-level execution gas,
(ii) intrinsic gas,
(iii) refunds, and
(iv) persistent state deltas, including net storage slots written and net contract bytecode growth.
This decomposition allows us to reason about execution as a multi-resource workload while remaining
grounded in the EVM's native accounting.

\item \emph{State sensitivity datasets.}
We build two complementary datasets that quantify dependence on historical state.
First, we probe the RPC surface by calling \texttt{eth\_estimateGas} at multiple historical states
and measuring how estimates and failure modes evolve as the estimation state drifts from the execution state.
Second, using replay-based tracing, we re-execute transactions on multiple historical states to
measure how opcode usage, storage access patterns, and total execution gas consumption change with state age.

\item \emph{Labeling and stratification.}
To connect low-level execution costs to higher-level behavior, we introduce transaction- and
address-level labels and use them to stratify both workload composition and state sensitivity across
application categories, including extractive, i.e., maximal extractable value (MEV), activity. State sensitivity is highly heterogeneous across transaction types.
\end{enumerate}

\subsection{Key Findings and Implications}

Our measurements reveal that workload shifts between L1 and L2 have direct implications not only for resource pricing and system efficiency, but also for workload estimation and transaction predictability experienced by users. We summarize the key takeaways below.

\begin{enumerate}
    \item \emph{One-dimensional pricing does not adapt to workload shifts.} The workload differences we observe between L1 and L2 are substantial enough that the right relative opcode prices should differ across chains. The instability is not confined to cross-chain differences: Ethereum's gas limit doubling during 2025 shifted its resource profile measurably toward more compute-heavy, Base-like patterns without any repricing of opcodes. When the workload mix drifts, one-dimensional gas metering cannot adapt on its own. The only recourse is manual repricing, which has historically lagged behind workload changes. In particular, Base L2 MEV is substantially different from Ethereum L1 MEV, being dominated by optimistic MEV~\cite{solmaz2025optimistic} (speculative on-chain probing) and more state-sensitive in both the variety of slots accessed and the values it writes. These findings motivate multi-dimensional fee markets pricing distinct resources separately.
    \item \emph{State growth is substantial (435 GB on Base, 30 GB on Ethereum in 2025), compositionally diverse, and bursty.} Storage slots account for 61.4\% of Base's growth and 50.9\% of Ethereum's, while bytecode contributes 25.4\% and 18.2\%, respectively. Simple transfers account for 33.2\% of Ethereum transactions and only 6.7\% of gas but contribute 15.4\% of state growth through persistent account creations. Daily state growth varies by up to an order of magnitude, and slot deletion rates are higher on Ethereum (34.6\%) than on Base (22.5\%), implying stronger incentives for storage reusability on Ethereum. Current gas pricing treats state writes as transient costs, failing to reflect their permanent nature; the true burden falls on every node.
    \item \emph{State sensitivity degrades transaction predictability and complicates execution design.} Gas estimates change across nearby historical states for 46.0\% of transactions on Base and 13.9\% on Ethereum, with MEV and DeFi activity exhibiting especially high sensitivity. Storage access patterns also diverge across execution states, limiting the effectiveness of access lists, complicating gas limit estimation for delayed execution designs, and increasing the cost of retries in optimistic parallel execution. This is already a user experience problem, and its consequences grow as chains adopt higher throughput, encrypted mempools, or delayed execution models.
\end{enumerate}

%% file: background.tex
\section{Background}
\label{sec:background}

\paragraph*{Transaction model and state machine replication.}

Blockchains follow the paradigm of state machine replication (SMR): at any given point, blockchain nodes maintain the state data structure; in Ethereum and EVM-compatible systems,
this state consists of accounts and their associated data (balances, contract code, and contract storage).
A distributed set of nodes then agree on an ordered log of transactions (consisting of an ordered chain
of blocks), and each node deterministically
applies every transaction in the order determined by the log to the previous maintained state
(i.e., we say that a transaction is executed against a particular \emph{pre-state}), in order
to reach the new state (we call this the \emph{post-state}) that will be used to apply the following
transaction in the ordered log, and so forth.

In practice, transactions are constructed and signed at a time during which both pending transactions for
inclusion at the following block might not be known, as well as the node or
entity (such as a wallet) might have an incomplete view of the possible pre-state of the transaction.
This can, e.g., happen in two key regimes: (1) when the chain is of high-throughput or low-latency and there are lots of
pending transactions or the transaction under-submission might only be included after an unpredictable
delay due to network factors, or (2) as an inherent
property of the system in the case of \emph{delayed execution}, i.e., an execution model in which consensus over transaction inclusion and ordering is reached before the corresponding state transitions have been computed, so that the last $k$ blocks remain unexecuted at submission time; examples include Monad~\cite{monadxyz2025} and Avalanche's ACP-194 specification~\cite{acp194}. Delayed execution increases performance by removing execution from the critical path of consensus, and pipelining it instead.
In any case, there is inherent uncertainty in the transaction's pre-state. As a result, the
same transaction's execution can follow different control-flow paths, touch different storage keys, or fail
when executed in the context of various possible pre-states. We say a transaction is \emph{state-sensitive} if re-executing it on a different pre-state produces different gas usage, different storage access patterns, or a different execution outcome (e.g., success vs.\ revert). The extent of this sensitivity is one of the key constituents under exploration in our work.

\paragraph*{Ethereum accounts, storage, and the EVM.}

Ethereum maps 160-bit addresses to \emph{accounts}. ``Externally owned accounts'' (EOAs in short) are
controlled by cryptographic private keys, and are transaction originators.
``Contract accounts'' additionally store EVM bytecode (corresponding to the contract) and maintain
persistent storage. Contract storage can be conceptually mapped to a key--value map from 256-bit
keys to 256-bit values. Storage slots can be accessed via the \texttt{SLOAD} (read) and \texttt{SSTORE}
(write) opcodes.

Persistent state is a long-term resource: it accumulates across blocks and directly affects the (ongoing)
storage as well as the sync costs of operating both a validator and a full node.
State can grow through the creation of new accounts (e.g., first-time recipients of value
transfers), new contract deployments (e.g., bytecode and initialized storage slots), and contract storage
writes.

The EVM is a stack-based virtual machine. Transactions can invoke smart contract functions directly or
indirectly via call-family opcodes (e.g., \texttt{CALL}, \texttt{DELEGATECALL}), which create nested call
frames and enable the composability of smart contract calls.
In addition to EVM bytecode, Ethereum exposes a small set of
special-purpose system contracts (called ``precompiles'') that implement, for instance, common
cryptographic and arithmetic operations at fixed addresses.

\paragraph*{Gas accounting and state-dependent costs.}

To guard against denial-of-service-style attacks, Ethereum charges transactions for each executed opcode in units of \emph{gas}, as outlined in \Cref{sec:intro}.
A transaction sender (hence the transaction itself) must specify a \emph{gas limit}, which caps the total gas that may be \emph{used} by execution. If execution runs out of gas, all state changes are reverted and the sender is charged the full gas limit of the failing frame. If the transaction reverts explicitly (e.g., via \texttt{REVERT}), state changes are also undone, but only the gas consumed up to the revert point is charged; the remainder is returned to the sender. Under Ethereum's EIP-1559 fee mechanism~\cite{eip1559}, blocks are constrained by a gas target and a maximum gas limit. The total gas used by a block is the sum of gas used by its included transactions in canonical order and must not exceed the block gas limit. The protocol adjusts a per-block \textit{base fee} in response to aggregate gas demand.
Users specify additional fee parameters (including a priority fee to be tipped to the validator of the
block), but the mapping from opcodes to gas costs is fixed by the protocol design.

Transaction gas can be further split into \emph{intrinsic} and \emph{execution} components.
Intrinsic gas covers fixed per-transaction overheads, including the base cost,
calldata charges, and (when present) up-front structural declarations
including, for example, access lists~\cite{eip2930} or authorization lists.
On the other hand, execution gas corresponds to the gas charged as per the EVM opcodes executed.
Ethereum also features gas refunds for certain operations (e.g., storage zeroing), but refunds are
bounded and have been reduced over time~\cite{eip3529} for reasons relating to the
reduction of DoS vectors.

State-dependence in gas costs arises from two mostly-disjoint mechanisms. First, \texttt{SSTORE} gas depends on the current slot value and whether allocation / deallocation occurs (zero $\leftrightarrow$ non-zero transitions) or merely overwrite occurs (non-zero $\leftrightarrow$ non-zero transitions), as codified by EIP-2200~\cite{eip2200} and refund rules under EIP-3529~\cite{eip3529}. Second, while EIP-2929~\cite{eip2929} cold/warm charges are local to a single transaction --- the first access within a transaction is always charged cold --- the \emph{set} of slots and accounts a transaction touches is itself state-dependent, because control flow and data-structure layout can lead the same transaction to access different keys under different pre-states. Access lists~\cite{eip2930} can pre-warm specified entries, but doing so effectively requires predicting this state-dependent access pattern in advance. In addition, the values stored in state can influence control flow, causing transactions to follow different execution paths depending on the state they encounter.

\paragraph*{Layer-2 solutions, typical applications, and workload types.}

EVM-compatible L2 solutions, called rollups, preserve EVM execution semantics while changing the
underlying execution regime. They typically offer much lower fees and higher throughput by executing transactions off-chain. Only minimally sufficient post-execution data is posted to the base layer (Ethereum L1 in this case) to substantiate the state transition.
Base, the L2 we study in this work, is an optimistic rollup. Ethereum L1 has $\sim$12-second blocks, whereas Base has $\sim$2-second blocks with higher gas capacity per block.

The core activity of the EVM lies in its smart contracts, which are particularly prolific
on high-throughput L2s. \emph{Tokens} (e.g., of the ERC-20 standard) implement
fungible balances maintained by accounts in persistent (contract) storage.
Many \emph{decentralized finance} (DeFi) applications, whose goal is to mirror functions of
traditional finance --- such as token exchange, lending, and liquidity provision --- often give rise to highly gas-
and storage-intensive contracts.
\emph{Maximal extractable value} (MEV) refers to profit opportunities obtained by
specialized actors from the strategic ordering, inclusion, and exclusion of transactions; one common MEV
strategy is exchange-rate arbitrage (for example between a \emph{centralized} and a
\emph{decentralized} exchange, otherwise known as CEX-DEX arbitrage).

%% file: related_work.tex
\section{Related Work}
\label{sec:related_work}

\paragraph*{Gas mispricing and the challenge of one-dimensional pricing.}
Prior work on gas pricing has focused on calibrating the relative pricing of opcodes to try to reflect the actual resource consumption within Ethereum L1. Perez and Livshits~\cite{perez2020broken} demonstrated how mispriced gas costs enable denial-of-service attacks. Parallel opcode-benchmarking efforts have measured the runtime costs of individual EVM opcodes under synthetic inputs to assess whether gas prices track real resource consumption~\cite{aldweesh2021opbench}; we complement this line of work by measuring the realized opcode mix in live 2025 workloads and how it drifts across chains and time.
Ethereum has undergone repeated gas cost adjustments (e.g., EIP-1884~\cite{eip1884}, EIP-2929~\cite{eip2929}).

However, even if these relative prices are well-calibrated, the one-dimensional model of metering leads to underutilization of some resources and overutilization of others when the workload mix changes~\cite{diamandis2023designing,angeris2024multidimensional}.
Our empirical findings in this work provide direct evidence that the mix of resource consumption varies both across time, and across different EVM ecosystems (L1 versus L2), thus motivating multidimensional fee markets. Beyond theoretical foundations, concrete proposals such as EIP-8011~\cite{eip8011} have begun to define multidimensional gas components for deployment, though they have not been grounded in an empirical study of live L1 and L2 workload mix drift.
Furthermore, simply inheriting L1 gas prices for L2 chains leads to resource imbalances due to differing workload characteristics.

Recent concurrent work by Silva~\cite{silva2025multidimensional} performs an empirical analysis of gas metering on Ethereum, breaking down gas usage according to opcodes and quantifying potential gains from multidimensional fee markets. While this work shares similarities with our gas decomposition methodology, we extend the analysis to compare L1 and L2 workloads, analyze the breakdown as a function of the activity type (such as DeFi, MEV, ERC20, etc.), and explicitly measure the sensitivity of the gas used by various opcodes to the state at which the transaction ends up executing.

\paragraph*{State growth and state access patterns.}
Silva~\cite{silva2025stategrowth} recently examined state growth scenarios and the impact of gas repricing on Ethereum, projecting future state sizes under different parameterizations. However, this analysis does not decompose state growth by transaction category or separately track storage slots versus contract bytecode. Our measurements quantify both dimensions and reveal that bytecode constitutes roughly 25\% of total state growth on Base, whereas simple transfers dominate state growth on Ethereum.
Ren et al.~\cite{ren2025analysis} analyze Ethereum workloads from a key-value storage perspective, focusing on storage access patterns and their implications for database design. Their work complements ours by examining storage system internals, whereas we focus on protocol-level resource consumption and its variation across execution contexts.
The persistent growth of blockchain state has motivated various proposals for state expiry and history pruning~\cite{buterin2024purge,buterin2021verkle,eip7927,eip4444}.
Our analysis highlights the current state of persistent state growth and shows, through transaction-level decomposition, how different application categories contribute to state bloat.

\paragraph*{Predictability of state access pattern for parallel and delayed execution.}
Many recent works~\cite{heimbach2023defi, wahrstaetter2025execution, biton2025ethereum} have analyzed state access patterns across transactions to understand parallelizability of EVM workloads. These works, however, assume that the set of storage slots accessed by a transaction is deterministic and known before execution. Our state sensitivity measurements challenge this assumption: we show that the same transaction can create different amounts of I/O workloads and access different storage keys when executed on different historical states. This variability threatens one of the key redeeming qualities of optimistic parallel execution strategies in the form of caching, as speculative execution may bring different data into cache than what is eventually needed upon retry after a conflict.

Heimbach et al.~\cite{heimbach2023dissecting} analyze EIP-2930 access lists and find that they are often populated incorrectly or suboptimally, resulting in missed gas savings or even gas penalties. Our work shows that optimal access lists are sensitive to the state at which the transaction ends up executing.
Cabral et al.~\cite{cabral2025demystification} investigate how gas usage and minimum gas limits vary across executions at different states, and the efficacy of different gas limit estimation algorithms as a function of the gap between estimation state and execution state. Our findings in this work are consistent with their results that the gas estimation quality degrades as the state keeps evolving.
We however take the sensitivity analysis further and break it down according to the sensitivity of various opcodes, and transaction categories. While Cabral et al. discover that largely gas estimates remain valid for about 11 blocks, we show that it depends on transaction type and blockchain context: e.g., MEV transactions on Base are far more state-sensitive than other types.

%% file: analysis.tex
\section{Data Collection}
\label{sec:data_collection}
Data is collected by running archival \texttt{reth} client Ethereum and Base nodes. While the analysis logic is client-agnostic, our data collection pipeline is implemented against \texttt{reth}-specific RPC interfaces,\footnote{\url{https://reth.rs/jsonrpc/trace}} as RPC call semantics and tracing implementations differ slightly across Ethereum client implementations. All code, scripts, queries, and the full opcode-to-category mapping used in this work are publicly available~\cite{evm_workload_code}.

\subsection{Transaction Gas Cost Decomposition and State Growth}\label{sec:longterm}
We now present the methodology for preparing our \textit{long-term dataset}.
For Ethereum and Base, we randomly sample 3{,}000 blocks per day for 2025. Sampling 3{,}000 blocks per day yields sufficient data for daily-granularity analysis. Note that processing all blocks, in combination with the multiple re-executions per transaction required for the state-sensitivity analysis, was computationally infeasible. This corresponds to sampling approximately 42\% of daily blocks on Ethereum (out of roughly 7{,}200 blocks per day) and approximately 7\% of daily blocks on Base (out of roughly 43{,}200 blocks per day). Blocks are sampled uniformly at random within each day and aggregate statistics such as state growth are extrapolated proportionally. While the sampling fraction is lower on Base, the absolute number of sampled transactions is comparable across chains (cf.\ Table~\ref{tab:gas_stats}).

For each sampled block, we decompose gas usage into intrinsic and execution components.
We then further break down both intrinsic gas and execution gas into their sub-components.
Since intrinsic gas is not exposed as a structured breakdown by the RPC interface, we reconstruct its components using explicit accounting rules.
Intrinsic gas comprises the minimum transaction gas (21,000), calldata costs, contract creation cost, access list gas (EIP-2930), and authorization list gas (EIP-7702). Execution gas is attributed per opcode. Each opcode is mapped to a single operation category (e.g., \texttt{SSTORE} to storage write, \texttt{SLOAD} to storage read); we do not decompose gas within an opcode. Gas refunds are tracked separately and capped according to EIP-3529. We additionally identify calls to EVM precompiles, including cryptographic and arithmetic precompiles, and account for protocol changes active during the measurement period, including the Pectra fork and EIP-7623 calldata pricing.

For the execution component, we trace all transactions and collect opcode-level execution data.
Each execution of a call-family opcode (\texttt{CALL}, \texttt{CALLCODE}, \texttt{DELEGATECALL}, and \texttt{STATICCALL}) creates a new call frame. In RPC execution traces, the gas attributed to a frame includes the gas charged by opcodes executed in that frame as well as the gas consumed by all nested child frames spawned by subsequent call opcodes. The gas consumed by the call-family opcode itself is not exposed as a separate quantity.

To calculate the gas charged for call-type opcodes, we apply a depth-directed accounting model. For each call frame, which is created by a call-family opcode, we compute the gas attributed to that call as the total gas reported for the frame minus the gas consumed by its direct child frames and minus the gas consumed by other opcodes executed directly in the same frame. This yields the gas charged for the call opcode itself. We note that our depth-directed accounting attributes gas to the frame that consumes it; gas that is reserved or forwarded by the protocol but never consumed is not counted as a cost in any frame.
We additionally track cold access patterns for account- and storage-level operations. This includes tracking account-related opcodes (e.g., \texttt{BALANCE}) and storage operations (\texttt{SLOAD}, \texttt{SSTORE}).

Finally, we quantify state growth over time. At the storage level, we track slot-level changes resulting from storage writes. We further record the creation of new accounts, as well as the deployment of contract code. These measurements allow us to track how storage and code footprint evolve over time.\footnote{Note that post-Cancun hard fork (EIP-6780), contract code removal is only possible for contracts created and destroyed within the same transaction.}

\subsection{State Sensitivity}
To characterize how sensitive transaction behavior is to the blockchain state, we perform a state sensitivity analysis that evaluates execution results, gas costs, and storage access patterns under different historical states. The analysis covers all transactions included in the 3{,}000 blocks sampled per day during September 2025, a period of stable protocol parameters between the Pectra fork (May 2025) and subsequent gas limit adjustments. Due to the computational cost of re-executing every transaction at multiple historical states, the analysis is limited to a single month.

We compare two kinds of execution states. The \emph{landed state} is the state under which a transaction actually executed on-chain, at its original position within block $b$. This serves as our ground-truth reference and is obtained from canonical execution traces via \texttt{debug\_\allowbreak traceBlockByNumber}. The \emph{lookback states} are historical pre-states at varying distance $N \geq 0$: for a transaction in block $b$, lookback $N$ refers to the state at the end of block $b - 1 - N$, so lookback $0$ is the state immediately preceding block $b$. Lookback states are obtained via \texttt{debug\_traceCall}. Gas estimation is relatively inexpensive compared to full opcode-level re-execution, which allows us to use finer lookback granularity ($N \in \{0, \ldots, 20\}$) for gas estimates while limiting gas decomposition and storage access analysis to a sparser set ($N \in \{0, 5, 10, 20\}$) plus the landed state.

\paragraph*{Gas estimate.} For each sampled block in the analysis period, we process all included transactions and evaluate gas estimates through the \texttt{eth\_estimateGas} API\footnote{\url{https://ethereum.org/developers/docs/apis/json-rpc/\#eth_estimategas}} at multiple historical states. Specifically, for a transaction included in block $b$, we request gas estimates at block heights $b - 1 - N$, with lookback distances $N \in \{0,1,2,\ldots,20\}$.

\paragraph*{Transaction gas cost decomposition.} We analyze state sensitivity by re-executing transactions on historical blockchain states. For a transaction originally included in block $b$, we replay execution at the landed state and lookback distances $N \in \{0, 5, 10, 20\}$. Re-execution for $N \geq 0$ is performed using \texttt{debug\_traceCall}, while the landed state uses canonical execution traces. Opcode-level gas usage is extracted as per the methodology of our long-term analysis (cf.\ Section~\ref{sec:longterm}). For each lookback level, we record execution success, total gas consumption, and per-opcode gas breakdowns.

\paragraph*{Storage slot access patterns.}
We collect the set of storage slots accessed by transactions at multiple state lookback points, comprising the landed state and lookback distances $N \in \{0, 5, 10, 20\}$. At each lookback level, we identify which storage slots are read and which are written during execution.\footnote{Write sets are obtained from state diffs (\texttt{prestateTracer} in diff mode) and thus record \emph{net-changed} slots, i.e., slots whose value after the transaction differs from their value before. Writes that restore a slot's original value, same-value writes, and writes reverted within the transaction are not captured.} We record both accessed slots and their values, enabling analysis of access-pattern and value divergence across historical states.

\subsection{Address and Transactions Labels}

We construct address- and transaction-level labels for Ethereum and Base using a multi-stage pipeline combining public labeling sources, on-chain measurements, and manual annotations. The pipeline is largely chain-agnostic, with chain-specific steps applied where required.

We first retrieve address labels from Spellbook (Dune Analytics~\cite{dune_spellbook}) and from DefiLlama~\cite{defillama_adapters}, which provides protocol metadata and categorical classifications across EVM-compatible chains. Next, we augment the label set with entries from the Kleros Curate registry~\cite{gmkung_blockscout_kleros_api}. Kleros Curate provides community-validated contract labels. These public sources cover 46.4\% of Ethereum transactions but only 19.3\% on Base, motivating the additional labeling steps described below, which raise coverage to 62.9\% and 75.8\%, respectively.

To identify protocol contracts that remain unlabeled
, we query the respective chain RPCs to detect common liquidity pools and token contracts, including Uniswap~V2/V3, Curve, Balancer, and Aerodrome deployments, as well as ERC20 tokens. This step is applied to all addresses with at least 100 transactions. We further enrich the labels using chain explorer APIs. For Ethereum, we query the Etherscan API, while for Base we query the Basescan API, targeting addresses with at least 1{,}000 transactions still unlabeled after earlier steps.

Next, we add chain-specific labels for MEV contracts. The labeling strategies differ between chains because the dominant MEV strategies differ: Ethereum L1 MEV is dominated by atomic arbitrage, sandwich, and CEX-DEX strategies, while on Base optimistic cyclic arbitrage dominates~\cite{solmaz2025optimistic}.
For Ethereum, we incorporate the top 1{,}000 addresses by gas usage, labeled as atomic arbitrage and CEX-DEX MEV contracts, obtained through Dune Analytics queries adapted from prior work~\cite{wu2025measuring,heimbach2024nonatomic,hildobby_mev_sandwich_trades,hildobby_atomic_mev_table}.
For Base, we add cyclic arbitrage MEV labels for the top 1{,}000 addresses by gas usage obtained through a Dune Analytics query adapted from prior work~\cite{solmaz2025optimistic}. For Base only, we further analyze unlabeled contracts with at least 10{,}000 transactions to assess whether they exhibit spam or optimistic MEV behavior. Optimistic MEV, i.e., speculative on-chain probing that frequently results in no state change, is a phenomenon specific to low-fee chains such as Base and is not prevalent on Ethereum L1~\cite{solmaz2025optimistic}. Thus, we apply this labeling step only to Base. For each selected contract, we sample up to 100 transactions and examine whether they induce any state changes beyond gas accounting.
We compute the proportion of sampled transactions that do not result in any state change other than gas accounting. Contracts for which more than half of the sampled transactions exhibit no state change are labeled as optimistic MEV. These MEV labels are applied only to addresses that remain otherwise unlabeled.

Finally, we manually label a small number of addresses that remain unlabeled after all automated steps. These labels are assigned in cases where external context is available, for example when a contract is deployed by a known deployer.

After label assignment, we map all address labels to a set of coarse-grained functional categories: \textit{MEV}, \textit{DeFi}, \textit{Token}, \textit{Infrastructure},\footnote{In the Infrastructure category, we group activity that supports the operation of on-chain protocols, such as price
oracle updates and account abstraction transactions (e.g., smart contract wallet execution and paymaster-sponsored gas).} \textit{Bridge}, \textit{NFT}, \textit{CEX}, and \textit{L2}.

In addition to address-level labels, we annotate transactions directly in the database. \textit{Simple ETH Transfers} are identified by a gas usage of exactly 21{,}000, excluding EIP-4844 blob transactions, which are labeled as \textit{L2}. \textit{Contract Creation} transactions are identified by a \texttt{NULL} receiver, that is, top-level deployments only. Transactions that merely execute the \texttt{CREATE} or \texttt{CREATE2} opcode, such as calls to factory contracts, retain their own category.

These transaction-level annotations take precedence over address-level labels: a transaction meeting the \textit{Simple ETH Transfer} criterion is counted as such even when its receiver carries a functional label.

\section{Execution and State Characteristics}
\label{sec:workload_main}

Our long-term dataset, built by sampling 3{,}000 blocks per day on both Ethereum and Base throughout 2025, comprises 219{,}811{,}453 Ethereum transactions and 256{,}506{,}972 Base transactions (cf.\ Table~\ref{tab:gas_stats}). At the transaction level, average gas usage is significantly higher on Base (232{,}936 gas per transaction) than on Ethereum (103{,}284 gas).

\subsection{Gas Decomposition}

\begin{table*}[t]
\vspace{-14pt}
\centering
\setlength{\tabcolsep}{4pt}
\begin{subtable}{0.52\textwidth}
\centering
\begin{tabular}{lrr}
\toprule
\textbf{Metric} & \textbf{Ethereum} & \textbf{Base} \\
\midrule
Total Transactions & 219,811,453 & 256,506,972 \\
\multicolumn{3}{l}{\textit{Mean Gas per Tx:}} \\
\quad Total Gas & 103,284 & 232,936 \\
\quad Execution & 82,441 & 217,340 \\
\quad Intrinsic & 26,656 & 25,654 \\
\quad Access List & 2,305 & 165 \\
\quad Auth List & 290 & 144 \\
\quad Refund & 6,770 & 10,735 \\
Execution \% of Total & 79.82\% & 93.30\% \\
\bottomrule
\end{tabular}
\caption{Per-Transaction Gas Statistics}\label{tab:gas_stats}

\end{subtable}
\hfill
\begin{subtable}{0.44\textwidth}
\centering
\begin{tabular}{lrr}
\toprule
\textbf{Category} & \textbf{Ethereum} & \textbf{Base} \\
\midrule
Storage Read & 22.73\% & 29.23\% \\
Storage Write & 34.88\% & 25.54\% \\
Compute & 20.21\% & 24.30\% \\
Calls & 10.18\% & 12.57\% \\
Contract Ops & 4.47\% & 4.78\% \\
Logging & 7.19\% & 3.45\% \\
Transient Storage & 0.15\% & 0.09\% \\
Epilogue & 0.20\% & 0.04\% \\
Precompiles & 0.00\% & 0.00\% \\
\bottomrule
\end{tabular}
\caption{Execution Gas Breakdown}\label{tab:exec_breakdown}
\end{subtable}\vspace{-3pt}

\caption{Gas usage comparison between Ethereum and Base in 2025. Table~\ref{tab:gas_stats} reports per-transaction gas statistics; access-list and authorization-list gas (EIP-7702) are included in intrinsic gas. Table~\ref{tab:exec_breakdown} reports the execution gas breakdown across nine EVM operation categories.}
\label{tab:gas_comparison}
\end{table*}

We commence the analysis by examining the execution gas decomposition in more detail (cf.\ Table~\ref{tab:gas_stats}). Notice that the execution component accounts for a substantially larger share of total gas usage on Base, comprising 93.30\% of total gas, compared to 79.82\% on Ethereum. This gap reflects both the larger share of simple value transfers on Ethereum, which incur only intrinsic gas (cf.\ Section~\ref{sec:address_category}), and the fact that the average transaction on Base uses more gas.

Breaking execution gas down by operation category (Table~\ref{tab:exec_breakdown}), we find that Base exhibits a more storage read-intensive workload than Ethereum, with storage reads accounting for 29.23\% of execution gas on Base compared to 22.73\% on Ethereum. In contrast, Ethereum spends a larger fraction of execution gas on storage writes. Base also allocates a higher share of execution gas to computation and call-related operations. Notably, a large fraction of call-related gas on Base is attributable to \texttt{STATICCALL}, which accounts for 66.8\% of call gas. In contrast, \texttt{STATICCALL} constitutes only 22.4\% of call-related gas on Ethereum. This indicates that cross-contract interactions on Base are predominantly read-only, reflecting a composability pattern centered on state inspection.

\begin{figure}[t]
\vspace{-10pt}
    \centering
    \begin{subfigure}{0.49\linewidth}
    \includegraphics[width=\linewidth]{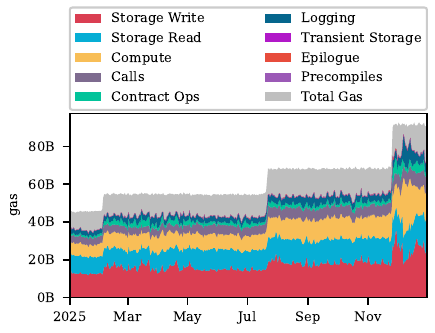}
    \caption{Ethereum}\label{fig:df_cat_daily_ethereum}
    \end{subfigure}\hfill
    \begin{subfigure}{0.49\linewidth}
    \includegraphics[width=\linewidth]{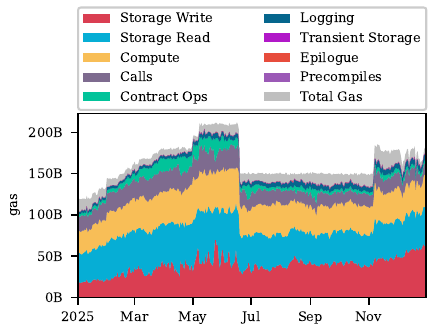}
    \caption{Base}\label{fig:df_cat_daily_base}
    \end{subfigure}
    \caption{Daily gas usage breakdown by EVM operation category for
    (\ref{fig:df_cat_daily_ethereum}) Ethereum and (\ref{fig:df_cat_daily_base}) Base.
    For Ethereum, we sample 3{,}000 blocks per day from approximately 7{,}200 daily blocks.
    For Base, there are 43{,}200 blocks per day, we likewise sample 3{,}000 blocks per day.
    Compared to Ethereum, Base shows a higher proportion of gas spent on storage reads and a lower share of non-execution overhead (shown in grey).
    }\label{fig:df_cat_daily}
\end{figure}

We further examine the temporal evolution of execution gas composition in Figure~\ref{fig:df_cat_daily} for Ethereum and Base. On Base, we observe that while the workload is initially dominated by storage reads, it gradually shifts toward a higher share of storage writes over time. This shift may be related to changes in MEV strategies on Base. In particular, optimistic MEV dominated gas usage in early 2025 and is characterized by extensive state probing via read-only interactions~\cite{solmaz2025optimistic}. Over time, we observe a reduction in such probing behavior and a corresponding increase in state-modifying execution (cf.\ Section~\ref{sec:address_category}), though other factors such as ecosystem maturation and gas limit increases during 2025 may also contribute to this shift. Other components such as logging remain comparatively stable.

Ethereum's execution mix also drifts measurably over the course of 2025: in January, storage writes accounted for 35.6\% of Ethereum's execution gas and compute for 18.0\%. By December, writes had eased to 32.5\% while compute rose to 23.8\%, even as total execution gas roughly doubled. This drift coincides with Ethereum's gas limit doubling from 30M to 60M during 2025, which increased capacity without repricing opcodes. The shift is in the direction of Base's more compute-heavy profile, although Ethereum remains more storage-write-intensive, suggesting that the capacity and cost regime a chain operates under shapes its workload composition.

\subsection{Storage Access Patterns}

To further characterize workload differences between Ethereum and Base, we examine storage access patterns by focusing on the \texttt{SLOAD} opcode, which reads values from contract storage. For each transaction, we record the total number of \texttt{SLOAD} operations and classify each access as cold or warm, depending on whether the corresponding storage slot is accessed for the first time in the transaction or has been previously accessed or pre-warmed via an access list.\footnote{An access list is a list of addresses and specific storage slots that a transaction declares in advance it intends to access.}
This distinction is important because warm storage accesses are charged significantly less gas than cold accesses. Under EIP-2929~\cite{eip2929}, the first \texttt{SLOAD} to a given storage slot in a transaction incurs a cold access cost of approximately 2{,}100 gas, whereas subsequent warm accesses to the same slot cost only about 100 gas.

\begin{table}[t]
\centering
\resizebox{\columnwidth}{!}{
\begin{tabular}{lrrrrrrrrrr}
\toprule
\textbf{Chain} & \textbf{Total Txs} & \textbf{Cold} & \textbf{Warm} & \multicolumn{6}{c}{\textbf{Accesses/Tx}} \\
\cmidrule(lr){5-10}
 & & & & \textbf{Mean} & \textbf{Std} & \textbf{P25} & \textbf{P50} & \textbf{P75} & \textbf{P95} \\
\midrule
Ethereum & 219,811,453 & 1,828,127,520 (39.6\%) & 2,791,584,524 (60.4\%) & 21.0 & 177.3 & 0.0 & 5.0 & 15.0 & 75.0 \\
Base & 256,506,972 & 7,401,539,024 (49.7\%) & 7,501,080,368 (50.3\%) & 58.1 & 261.1 & 3.0 & 14.0 & 49.0 & 243.0 \\
\bottomrule
\end{tabular}}
\caption{\texttt{SLOAD} access pattern comparison between Base and Ethereum, showing total transactions, cold versus warm storage reads, and per-transaction \texttt{SLOAD} access statistics.
}\label{tab:sload}
\end{table}

Table~\ref{tab:sload} summarizes the results. Base transactions access more distinct storage slots than Ethereum, indicating lower key reuse: 49.7\% of \texttt{SLOAD} operations on Base are cold, vs.\ 39.6\% on Ethereum.

Base transactions also execute substantially more \texttt{SLOAD}s on average than Ethereum, with a heavier tail and a P25 of zero on Ethereum reflecting many transactions with no storage interaction (cf.\ Table~\ref{tab:sload}).

Cold storage accesses are generally considered underpriced relative to other opcodes, even after the adjustments introduced through EIP-2929. Recent protocol designs and proposals, including newer EVM variants such as Monad~\cite{categorylabs_monad_initial_spec_proposal} and ongoing discussions within the Ethereum Foundation~\cite{eip8038}, have introduced or considered higher costs for cold storage reads. In this context, the higher fraction of cold \texttt{SLOAD} operations and the larger number of distinct storage accesses observed on Base indicate that a larger share of the workload is concentrated in components that are comparatively underpriced.

\subsection{State Growth}

Related to cold storage accesses, state growth is commonly cited as a key constraint for scaling blockchain systems. We therefore continue with an analysis of state growth. Table~\ref{tab:storage_growth} reports estimated yearly state growth on Ethereum and Base, with figures extrapolated from our sampled dataset. We extrapolate by dividing sampled counts by the per-chain sampling fraction (42\% for Ethereum, 6.9\% for Base). Thus, cross-chain ratios compare extrapolated full-year totals, not raw sampled counts.

We decompose state growth into three components: newly added storage slots, deployed smart contract bytecode, and account state. Storage growth from slots is computed assuming 64 bytes (i.e., 32 bytes for the key and value respectively). Account state includes both EOAs and contract accounts and consists of four 32-byte fields: nonce, balance, code hash, and storage root. This results in 128 bytes per account. Note that the exact byte footprint of account state may vary across client implementations due to differences in data representation. Our approximation provides a consistent basis for comparison.

\begin{table*}[t!]
\vspace{-10pt}
\centering

\begin{subtable}{0.48\textwidth}
\centering
\resizebox{\textwidth}{!}{
\begin{tabular}{lrr}
\toprule
\textbf{Metric} & \textbf{Ethereum} & \textbf{Base} \\
\midrule
\multicolumn{3}{l}{\textit{Storage Slots:}} \\
\quad Created & 364,008,598 & 5,388,214,765 \\
\quad Deleted & 125,870,269 & 1,209,883,109 \\
\quad Updated & 1,254,823,628 & 12,699,968,280 \\
\quad Net Written & 238,138,180 & 4,178,331,508 \\
\midrule
\multicolumn{3}{l}{\textit{Contract Bytecode (GB):}} \\
\quad Allocated & 5.44 & 110.60 \\
\quad Freed & 0.00 & 0.00 \\
\quad Net Growth & 5.44 & 110.60 \\
\midrule
\multicolumn{3}{l}{\textit{Accounts:}} \\
\quad Created & 72,230,074 & 446,944,680 \\
\quad Deleted & 33,653 & 4,905 \\
\quad Net Growth & 72,196,421 & 446,939,775 \\
\bottomrule
\end{tabular}
}
\caption{State Changes}\label{tab:storage_changes}
\end{subtable}
\hfill
\begin{subtable}{0.48\textwidth}
\centering
\resizebox{\textwidth}{!}{
\begin{tabular}{lrr}
\toprule
\textbf{Metric} & \textbf{Ethereum} & \textbf{Base} \\
\midrule
\multicolumn{3}{l}{\textit{Total Storage Growth (GB):}} \\
\quad Storage Slots & 15.24 & 267.41 \\
\quad Bytecode & 5.44 & 110.60 \\
\quad Account State & 9.24 & 57.21 \\
\quad \textbf{Total} & \textbf{29.92} & \textbf{435.22} \\
\midrule
\multicolumn{3}{l}{\textit{Per-Transaction Averages:}} \\
\quad Slots/TX & 0.45 & 1.13 \\
\quad Bytecode/TX (bytes) & 10.3 & 29.9 \\
\midrule
\multicolumn{3}{l}{\textit{Ratios (Base/Ethereum):}} \\
\quad Net Slots & \multicolumn{2}{r}{17.55x} \\
\quad Bytecode Growth & \multicolumn{2}{r}{20.33x} \\
\quad Net Accounts & \multicolumn{2}{r}{6.19x} \\
\quad Total Storage & \multicolumn{2}{r}{14.55x} \\
\bottomrule
\end{tabular}
}
\caption{Storage Totals \& Ratios}\label{tab:storage_totals}
\end{subtable}

\caption{State storage growth on Ethereum and Base in 2025, extrapolated from the sampled blocks. Table~\ref{tab:storage_changes} reports storage-slot creations, deletions, and updates; net slot growth; bytecode allocation and removal; and account creations and deletions.\protect\footnotemark{} Table~\ref{tab:storage_totals} converts net growth to logical GB using 64 bytes per storage slot and 128 bytes per account, reports per-transaction averages over all transactions, and gives Base/Ethereum ratios.}
\label{tab:storage_growth}
\end{table*}
\footnotetext{Under EIP-161~\cite{eip161}, an account is dead if it is non-existent or empty, i.e., it has no code, zero nonce, and zero balance. We count dead-to-live and live-to-dead transitions at transaction boundaries, and net account growth is the difference between the two counts. A live-to-dead transition occurs when a prefunded address with zero nonce and no code receives a \texttt{CREATE2} deployment, which EIP-684~\cite{eip684} permits because balance is not a collision criterion, and the resulting contract executes \texttt{SELFDESTRUCT} in the same transaction, for which EIP-6780~\cite{eip6780} still deletes the account. The prefunding is what makes the account live before the transaction, so the deletion is visible at a transaction boundary.}

In 2025, estimated state growth differs markedly between Base and Ethereum. Based on extrapolation from our sampled data, total state growth amounts to 435.22 GB on Base, compared to 29.92 GB on Ethereum, corresponding to a 14.55$\times$ difference. A gap in this direction is expected given Base's $\sim$6$\times$ higher block rate, substantially lower fees, and greater share of storage-write-heavy activity. The observed magnitude, however, exceeds a naive throughput-based extrapolation, indicating that workload composition, not just transaction volume, drives the difference.

The composition of this growth also differs substantially. On Base, newly added storage slots account for 61.4\% of total state growth, compared to 50.9\% on Ethereum, while deployed contract bytecode constitutes 25.4\% on Base and 18.2\% on Ethereum. Account state represents 30.9\% of total state growth on Ethereum, compared to 13.1\% on Base.

Account state growth also differs vastly between the two chains. On a per-transaction basis, Base transactions induce significantly more persistent state, with an average net growth of 1.13 storage slots per transaction compared to 0.45 on Ethereum, and 29.9 bytes of net bytecode growth per transaction compared to 10.3 bytes on Ethereum. This is reflective of the difference in dollar cost between the two chains, along with the usage patterns and maturity of the ecosystem. A larger fraction of newly created storage slots is subsequently deleted on Ethereum (approximately 35\%) than on Base (approximately 22\%), indicating higher slot turnover and stronger incentives for storage reuse on Ethereum.

To illustrate how individual contracts contribute to state growth, we examine the address \href{https://basescan.org/address/0x9ec1c3dcf667f2035fb4cd2eb42a1566fd54d2b7}{\texttt{0x9ec...2b7}}. Within our sampled blocks, this address is involved in 86{,}046 transactions and contributes 336{,}378{,}240 bytes of deployed contract bytecode, 37{,}315{,}630 net added storage slots (37{,}534{,}300 created), and 7{,}475{,}172 account births. In aggregate, this corresponds to an estimated 3.681~GB of additional state in our sampled blocks, which extrapolates to 53.01~GB of yearly storage added. This address acts as a deployer for the XEN token, which is known to generate substantial on-chain storage usage~\cite{xen_token_state_bloat}, and serves as an example of how a single contract can account for a non-negligible share of observed state growth. This single contract represents 12.18\% of storage growth while only making up 0.03\% of transactions.

In addition to storage being considered one of the main bottlenecks for blockchain scalability~\cite{halborn_blockchain_storage_problem}, the EVM collapses a high-dimensional resource-consuming workload into a single scalar unit of metering (gas). As a result, long-term state growth is priced indirectly through short-term, per-block execution limits, creating unnecessary fluctuations in the price of state growth. To examine the implications of this design choice, we analyze the distribution of daily state growth on Ethereum and Base.

Daily storage growth varies substantially over time on both chains (cf.\ Table~\ref{tab:storage_distribution} in Appendix~\ref{app:storage_distribution}). For example, on Base, daily storage growth ranges from 354.5~MB at the 1st percentile to 3{,}075.3~MB at the 99th percentile. This variance is driven by bursty, ecosystem-specific events, with concentration from a small number of contracts often dominating daily state totals.

The large variance in daily state growth highlights a fundamental misalignment between short-term, transient execution-based gas pricing and the long-term, cumulative costs of persistent storage. Per-block gas limits can only regulate instantaneous resource usage, forcing a trade-off between restrictive limits that cause overall underutilization of chain and permissive limits that allow bursts of activity to translate into excessive and persistent state growth. As a result, volatile workloads are either constrained in the short term or, in the worst case, allowed to cause excessive long-term storage growth under the current gas pricing.

\subsection{Behavior by Address Category}\label{sec:address_category}

\begin{figure}[t!]\vspace{-10pt}
    \centering
    \begin{subfigure}{0.49\linewidth}
    \includegraphics[scale=0.93]{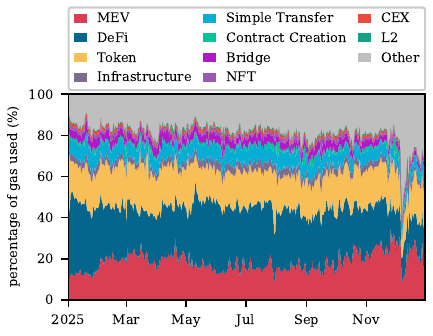}
    \caption{Ethereum}\label{fig:tx_category_gas_percentage_ethereum}
    \end{subfigure}\hfill
    \begin{subfigure}{0.49\linewidth}
    \includegraphics[scale=0.93]{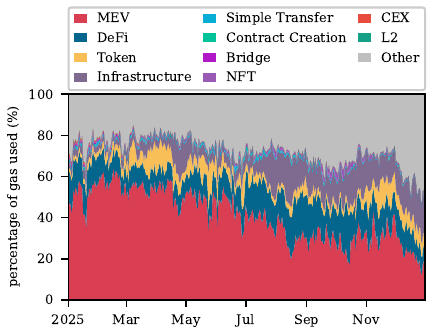}
    \caption{Base}\label{fig:tx_category_gas_percentage_base}
    \end{subfigure}
    \caption{Daily share of total gas usage by transaction category on (\ref{fig:tx_category_gas_percentage_ethereum}) Ethereum and (\ref{fig:tx_category_gas_percentage_base})  Base over 2025.
    Gas usage is normalized per day and aggregated by transaction category. The anomaly in the share of gas used visible on Ethereum in December 2025 coincides with a \texttt{Prysm} client issue immediately following the Fusaka hard fork.
    }\label{fig:tx_category_gas_percentage}
\end{figure}

Next, we investigate differences in workload composition by transaction category, considering both gas usage and storage growth. We focus on the main categories observed on both chains: MEV, DeFi applications such as DEXs and lending protocols, ERC-20 token interactions, infrastructure services such as oracles or account abstraction, and simple transfers. On Base, transaction activity is also more concentrated across a smaller set of receiver addresses, reflecting a more clustered workload structure (cf.\ Appendix~\ref{app:address}). Figure~\ref{fig:tx_category_gas_percentage} shows the evolution of gas usage by category over time, while Table~\ref{tab:gas_storage_by_category} summarizes transaction counts, average gas usage, and average storage growth per transaction.

MEV constitutes a particularly large share of activity on Base, accounting for 41.2\% of total gas usage and 43.6\% of transactions. Although its share decreases toward the end of the year (Figure~\ref{fig:tx_category_gas_percentage_base}), MEV remains the dominant gas-consuming category. In contrast, MEV accounts for only 18.8\% of total gas usage on Ethereum. Notice that towards the end of the year, the ``Other'' category grows on Base, suggesting that some emerging or evolving MEV patterns may not be fully captured by our conservative labeling (i.e., only labeling addresses with a significant number of transactions and gas usage). On Ethereum, ERC-20 token transactions account for 29.1\% of transactions and 17.9\% of total gas usage, whereas on Base they account for only 7.9\% of transactions and 6.0\% of gas usage. DeFi activity contributes a similar share of gas usage on both chains, although DeFi transactions on Base consume more gas per transaction on average.

Appendix~\ref{app:gas_by_category} provides a detailed gas decomposition by category. For MEV transactions, the breakdown shows a more storage read-intensive and less storage write-intensive profile on Base than on Ethereum, consistent with optimistic MEV that reads on-chain prices during execution. Infrastructure transactions on Base, by contrast, exhibit a strongly storage write-heavy profile. Overall, the decomposition reveals systematic differences in workload composition across categories and chains.

\begin{table*}[t!]\vspace{-2pt}
\centering
\resizebox{\textwidth}{!}{
\begin{tabular}{lrrrrrrrrrrrr}
\toprule
\textbf{Category} & \multicolumn{6}{c}{\textbf{Ethereum}} & \multicolumn{6}{c}{\textbf{Base}} \\
\cmidrule(lr){2-7} \cmidrule(lr){8-13}
& \multicolumn{2}{c}{\textbf{Transactions}} & \multicolumn{2}{c}{\textbf{Gas Used}} & \multicolumn{2}{c}{\textbf{Storage Growth}} & \multicolumn{2}{c}{\textbf{Txs}} & \multicolumn{2}{c}{\textbf{Gas Used}} & \multicolumn{2}{c}{\textbf{Storage Growth}} \\
\cmidrule(lr){2-3} \cmidrule(lr){4-5} \cmidrule(lr){6-7} \cmidrule(lr){8-9} \cmidrule(lr){10-11} \cmidrule(lr){12-13}
& \textbf{\#} & \textbf{Pct} & \textbf{Avg} & \textbf{Pct} & \textbf{Avg (KB)} & \textbf{Pct} & \textbf{\#} & \textbf{Pct} & \textbf{Avg} & \textbf{Pct} & \textbf{Avg (KB)} & \textbf{Pct} \\
\midrule
MEV & 14,458,276 & 6.6 & 295,627 & 18.8 & 0.17 & 20.1 & 111,896,672 & 43.6 & 220,110 & 41.2 & 0.04 & 14.8 \\
DeFi & 31,854,577 & 14.5 & 180,632 & 25.3 & 0.03 & 7.7 & 33,704,931 & 13.1 & 228,276 & 12.9 & 0.06 & 6.3 \\
Token & 63,950,730 & 29.1 & 63,563 & 17.9 & 0.03 & 14.6 & 20,354,249 & 7.9 & 174,677 & 6.0 & 0.22 & 15.0 \\
Infrastructure & 3,382,961 & 1.5 & 210,106 & 3.1 & 0.07 & 2.0 & 7,368,706 & 2.9 & 792,576 & 9.8 & 1.05 & 25.6 \\
Simple Transfer & 72,928,518 & 33.2 & 21,000 & 6.7 & 0.03 & 15.4 & 20,152,658 & 7.9 & 21,000 & 0.7 & 0.02 & 1.2 \\
Bridge & 5,769,963 & 2.6 & 138,939 & 3.5 & 0.04 & 1.7 & 1,779,509 & 0.7 & 140,766 & 0.4 & 0.02 & 0.1 \\
NFT & 2,363,354 & 1.1 & 140,973 & 1.5 & 0.08 & 1.5 & 2,770,415 & 1.1 & 134,890 & 0.6 & 0.08 & 0.7 \\
Contract Creation & 215,137 & 0.1 & 1,670,843 & 1.6 & 5.52 & 9.5 & 214,308 & 0.1 & 1,128,603 & 0.4 & 4.65 & 3.3 \\
CEX & 1,760,373 & 0.8 & 169,362 & 1.3 & 0.18 & 2.6 & 256,270 & 0.1 & 265,441 & 0.1 & 0.05 & 0.0 \\
L2 & 1,525,605 & 0.7 & 74,233 & 0.5 & 0.03 & 0.3 & 0 & 0.0 & - & 0.0 & - & 0.0 \\
Other & 21,601,959 & 9.8 & 206,551 & 19.7 & 0.14 & 24.4 & 58,009,254 & 22.6 & 287,422 & 27.9 & 0.17 & 32.9 \\
\bottomrule
\end{tabular}}
\caption{Gas usage and storage growth by address category on Ethereum and Base in 2025. For each chain, shows transaction count and percentage, average gas per transaction and percentage of total gas, average storage per transaction (KB) and percentage of total storage. "Other" includes unlabeled transactions and categories with less than 0.1\% gas usage on both chains.}
\label{tab:gas_storage_by_category}
\end{table*}

Storage growth further differentiates transaction categories. Contract creation dominates per-transaction storage growth on both chains, which is not
surprising given that transactions in this category deploy new contracts. Infrastructure transactions on Base also contribute disproportionately to overall storage growth relative to their share of transactions. In contrast, simple transfers account for a large fraction of transactions on Ethereum (33.2\%), while contributing little to total gas usage (6.7\%) yet 15.4\% of state growth.
This pattern indicates a high incidence of simple transfers to new addresses on Ethereum, increasing account-state growth. Notably, this form of state expansion is not directly reflected in transaction fees, as simple transfers incur only the minimum intrinsic gas cost of 21{,}000 gas.

\section{State Sensitivity Analysis}

In addition to analyzing how transactions behave in the specific block and position in which they are included, we also study how their behavior varies as a function of the execution state.
This perspective is particularly relevant for user experience on high-throughput chains, where multiple blocks may be processed between transaction creation by the user and execution, allowing the underlying state to change. It is likewise important in blockchains with delayed execution~\cite{monadxyz2025,acp194}, where the execution of preceding blocks may not have completed at submission time and the exact state against which a transaction executes is not yet known.

To capture these effects, we execute each transaction we sampled in September 2025 against historical states at varying lookback distances $N$, emulating the view of the user creating the transaction. Specifically, a transaction included in block $b$ is re-executed on the state corresponding to the end of block $b-1-N$.

\subsection{Gas Estimation}

\begin{table}[t]
\centering
\resizebox{\textwidth}{!}{
\begin{tabular}{lrrrrrrrrrrrrrrr}
\toprule
& & & \multicolumn{6}{c}{\textbf{CV (\%)}} & \multicolumn{5}{c}{\textbf{Max Successful Simulation Lookback}} \\
\cmidrule(lr){4-9} \cmidrule(lr){10-14}
\textbf{Chain} & \textbf{Txs} & \textbf{Non-Zero (\%)} & \textbf{Mean} & \textbf{P50} & \textbf{P75} & \textbf{P90} & \textbf{P95} & \textbf{P99} & \textbf{P1} & \textbf{P5} & \textbf{P10} & \textbf{P25} & \textbf{P50} \\
\midrule
Ethereum & 19,262,259 & 13.9 & 0.57 & 0.00 & 0.00 & 0.41 & 2.79 & 14.84 & 0 & 1 & 3 & 15 & 20 \\
Base & 22,420,978 & 46.0 & 6.88 & 0.00 & 0.91 & 15.65 & 40.64 & 139.74 & 0 & 4 & 15 & 20 & 20 \\
\bottomrule
\end{tabular}
}

\caption{Gas estimate variance across state lookback distances for September 2025.
CV denotes the coefficient of variation (standard deviation divided by mean) of gas estimates across lookbacks.
Max Successful Simulation Lookback reports the maximum lookback distance (in blocks), up to 20, at which transaction simulation succeeds without error.}
\label{tab:estimate_variance_summary}
\end{table}

We begin by analyzing the values returned by the gas estimation API, \texttt{eth\_estimateGas}, which estimates the minimum gas required for a transaction's successful execution, across a range of state lookback distances. For each transaction, we simulate execution on historical states up to a maximum lookback of 20 blocks. If execution fails at a given lookback distance $x$, we do not continue simulations further into the past. For transactions that execute successfully across multiple lookbacks, we record (i) the coefficient of variation (CV) of the gas estimate, and (ii) the maximum lookback distance at which execution succeeds. We define the CV of a transaction's gas estimate as $\mathrm{CV} = \sigma / \mu$, where $\mu$ and $\sigma$ are the mean and standard deviation of the gas estimates obtained across successful lookbacks. A CV of zero indicates that the gas estimate is identical across all lookback distances, while a higher CV indicates greater sensitivity to execution state. The reported transaction counts and non-zero fractions cover all transactions with at least one successful estimate. The CV itself is defined only for transactions with at least two successful estimates, which make up 95.1\% of the population on Ethereum and 98.8\% on Base.

On Ethereum, only 13.9\% of the full population have observed non-zero gas estimate variance across lookbacks, compared to 46.0\% on Base (cf.\ Table~\ref{tab:estimate_variance_summary}). Consistent with this, the mean coefficient of variation (CV) of gas estimates is substantially higher on Base than on Ethereum. This difference persists across the distribution, as reflected in consistently higher percentile values (e.g., 14.84\% at the 99th percentile on Ethereum versus 139.74\% on Base). Overall, this indicates greater sensitivity of gas requirements to execution state on Base than on Ethereum. This difference likely reflects higher per-block transaction throughput and thereby more frequent state updates per block on Base along with a different workload composition. We note that our lookback windows are defined in blocks rather than wall-clock time. Since Ethereum produces blocks every ${\sim}12$ seconds and Base every ${\sim}2$ seconds, 20 blocks corresponds to ${\sim}240$ seconds on Ethereum but only ${\sim}40$ seconds on Base. The closest comparable wall-clock interval on Ethereum is a lookback of 5 blocks (${\sim}60$ seconds), which still exceeds the Base window. Even under this more conservative comparison, the sensitivity gap remains substantial, indicating that Base experiences genuinely higher state churn once block-time differences are accounted for.

The distribution of maximum successful simulation lookback distances, on the other hand, shows that simulations on Base tend to remain successful further into the past than on Ethereum. In particular, the 25th percentile is 20 blocks on Base, compared to 15 blocks on Ethereum. This is not a pure measure of code executability: because simulations retain the original transaction fee fields, a transaction can be rejected when its fee cap is below the historical base fee. The cross-chain difference may therefore reflect block times and fee dynamics as well as state-dependent execution.\footnote{The minimum fee a transaction needs to pay to be included in a particular block.}

We further examine gas estimate variance by transaction category (cf.\ Table~\ref{tab:estimate_variance_by_category} in Appendix~\ref{app:estimate_variance_category}). MEV and DeFi transactions affect the largest share of transactions with non-zero gas estimate variance on both chains (54.2\% and 53.0\% on Base), and MEV also carries the largest CV magnitude there (mean CV 10.11\%). Contract Creation is instead among the least variable categories on Base (0.0\% non-zero variance), since a deployment's gas cost is dominated by its own init code rather than by the state it executes against. Notably, the maximum successful lookback for MEV transactions on Base is substantially larger than on Ethereum (5th percentile at 20 blocks on Base versus 0 on Ethereum), likely reflecting the contrast between traditional MEV strategies on L1 and optimistic MEV strategies on L2s, which are by design crafted to operate under unknown state.

At the low-variance end, simple transfers exhibit greater state sensitivity on Ethereum (3.0\% non-zero variance) than on Base (0.0\%). This variance arises from EIP-7702, which allows EOAs to temporarily delegate execution to contract code, causing the same transfer to execute differently depending on whether delegation was active at a given historical state.

Overall, DeFi and MEV transactions account for the largest share of transactions whose gas requirements vary with the underlying execution state, while variability on Ethereum is also driven by time-related effects.

\subsection{Gas Decomposition}

Next, we analyze gas decomposition across state lookbacks to identify which gas components exhibit the highest variability, including storage operations, logging, and computation, as well as variability in total transaction gas. This analysis includes transactions with at least two successful executions across the considered lookback distances, and the variability metrics are computed over the successful executions of each transaction. Transactions that succeed at fewer than two lookbacks are excluded, and failed executions do not contribute to the reported variability. Most included transactions succeed at all four lookbacks (91.1\% on Base and 85.6\% on Ethereum), with the remainder contributing two or three observations. We evaluate execution at lookback distances of 0, 5, 10, and 20 blocks.

\begin{table}[!t]\vspace{-10pt}
\centering
\resizebox{\textwidth}{!}{
\begin{tabular}{lrrrrrrrr}
\toprule
& \textbf{Overall} & \textbf{S-Read} & \textbf{S-Write} & \textbf{Trans} & \textbf{Compute} & \textbf{Calls} & \textbf{Contract} & \textbf{Log} \\
\midrule
\textbf{Ethereum CV\%} & 0.64 & 0.56 & 1.55 & 0.08 & 0.63 & 0.60 & 0.50 & 1.03 \\
\textbf{Ethereum \% NZ} & 9.5 & 9.4 & 9.5 & 2.3 & 9.5 & 8.8 & 7.9 & 9.4 \\
\midrule
\textbf{Base CV\%} & 5.22 & 2.82 & 6.76 & 1.64 & 3.20 & 3.70 & 7.71 & 4.40 \\
\textbf{Base \% NZ} & 42.6 & 42.5 & 20.1 & 5.7 & 42.6 & 42.1 & 20.7 & 18.1 \\
\bottomrule
\end{tabular}
}
\vspace{1mm}
\caption{Gas variance by opcode group across state lookback distances. The table reports the mean coefficient of variation (CV\%) and the fraction of transactions with non-zero (NZ) variance. The overall columns refer to total transaction gas. The opcode-group columns refer to each group's share of total transaction gas, capturing shifts in gas composition across execution states.}
\label{tab:gas_variance_opcode_group}
\end{table}

\vspace{10pt}

Table~\ref{tab:gas_variance_opcode_group} shows that gas variability is higher on Base than on Ethereum, with overall gas CV roughly $8\times$ higher on Base (5.22\% versus 0.64\%) and larger shifts in the composition of gas across opcode groups.\footnote{This analysis uses four replay states, whereas gas estimation uses 21, and therefore has fewer opportunities to detect variation. Because variability is computed only across successful re-executions, the reported CVs and non-zero shares exclude state changes that cause replay failure and should be interpreted as conservative estimates.}

On Ethereum, the compositional variability is concentrated in storage writes and contract-related operations. On Base, it spans storage reads, writes, compute, and calls (cf.\ Table~\ref{tab:gas_variance_opcode_group}).

The per-category breakdown (cf.\ Table~\ref{tab:gas_variance_category_opcode} in Appendix~\ref{app:gas_variance_category}) reveals that MEV transactions on Base exhibit the highest variability of any Base category, with the mean CV of the opcode group's gas share reaching 11.45\% for contract-related opcodes, 9.40\% for storage writes, and 5.55\% for logging. Base \textit{Infrastructure} follows a similar pattern at a slightly lower level (9.11\% and 8.21\% respectively), while DeFi is more moderate (up to 4.33\%). On Ethereum most categories are considerably less variable, but \textit{Infrastructure} is the exception and carries the single largest value in the table, a 15.91\% mean share CV for storage writes. Variability is thus concentrated in the categories whose gas composition depends most on the state they encounter.

\subsection{Storage Slot Access Patterns}

Finally, we analyze variance in storage slot accesses across state lookbacks. Specifically, we examine whether the set of storage slots accessed by a transaction changes as a function of the execution state. High variability in accessed slots has important implications. First, it limits the effectiveness of access lists, which require transactions to predeclare accessed storage slots in order to receive a gas discount. If accessed slots vary across execution states, access lists cannot be populated reliably. Second, many optimistic parallel execution approaches execute transactions speculatively in parallel and roll back upon conflicts, assuming that re-execution is faster due to cached state. This is no longer the case if retrying execution on a different state results in accesses to uncached storage slots.

We quantify storage slot overlap between two lookback distances $N_1$ and $N_2$ using three per-transaction metrics based on the Jaccard index. Let $R_i$ and $W_i$ denote the sets of storage slots read and written, respectively, at lookback $N_i$. \emph{Read overlap} is the Jaccard similarity of the readsets, i.e., $|R_1 \cap R_2| \,/\, |R_1 \cup R_2|$. \emph{Write overlap} is the Jaccard similarity of the writesets, i.e., $|W_1 \cap W_2| \,/\, |W_1 \cup W_2|$. \emph{Value consistency} is the fraction of slots in the write intersection whose written values are identical, i.e., $|\{s \in W_1 \cap W_2 : v_1(s) = v_2(s)\}| \,/\, |W_1 \cap W_2|$. Each metric is computed over the transactions for which it is defined, so the three populations differ. Read overlap requires non-empty readsets at both lookbacks, which also excludes transactions whose re-execution produced no trace at either lookback. Write overlap additionally requires a non-empty write union. A transaction whose writeset vanishes entirely at one lookback scores zero, while transactions with empty writesets at both lookbacks are excluded. Value consistency requires an overlapping write. On Base at the (landed, $N=0$) pair, these populations comprise 23.3M, 10.1M, and 8.6M transactions, respectively. We report the mean and median of each metric across its eligible transactions.

Appendix~\ref{app:slots-per-tx} reports the distribution of the number of storage slots accessed per transaction at each lookback distance. We find that Base transactions access substantially more slots on average (22.0 at the landed state) than Ethereum (14.7), with high variability. Overlap across lookbacks is also limited, particularly on Base.

\begin{table}[t!]\vspace{-10pt}
\centering

\resizebox{\textwidth}{!}{
\begin{tabular}{lrrrrrrrrrrrr}
\toprule
& \multicolumn{6}{c}{\textbf{Ethereum}} & \multicolumn{6}{c}{\textbf{Base}} \\
\cmidrule(lr){2-7} \cmidrule(lr){8-13}
\textbf{Lookback Pair} & \multicolumn{2}{c}{\textbf{Read (\%)}} & \multicolumn{2}{c}{\textbf{Write (\%)}} & \multicolumn{2}{c}{\textbf{Value (\%)}} & \multicolumn{2}{c}{\textbf{Read (\%)}} & \multicolumn{2}{c}{\textbf{Write (\%)}} & \multicolumn{2}{c}{\textbf{Value (\%)}} \\
\cmidrule(lr){2-3} \cmidrule(lr){4-5} \cmidrule(lr){6-7} \cmidrule(lr){8-9} \cmidrule(lr){10-11} \cmidrule(lr){12-13}
 & \textbf{Mean} & \textbf{Median} & \textbf{Mean} & \textbf{Median} & \textbf{Mean} & \textbf{Median} & \textbf{Mean} & \textbf{Median} & \textbf{Mean} & \textbf{Median} & \textbf{Mean} & \textbf{Median} \\
\midrule
Landed vs 0 &  97.1 & 100 &  94.6 & 100 &  78.5 & 100 &  93.5 & 100 &  80.8 & 100 &  67.8 &  80 \\
Landed vs 5 &  92.2 & 100 &  83.5 & 100 &  65.3 &  83 &  87.9 & 100 &  66.5 & 100 &  50.6 &  50 \\
Landed vs 10 &  90.8 & 100 &  79.7 & 100 &  62.4 &  67 &  86.4 & 100 &  63.0 &  98 &  46.7 &  50 \\
Landed vs 20 &  89.3 & 100 &  75.6 & 100 &  59.6 &  50 &  84.9 & 100 &  59.3 &  86 &  43.1 &  36 \\
 0 vs 5 &  94.3 & 100 &  86.7 & 100 &  68.7 & 100 &  90.0 & 100 &  72.6 & 100 &  52.0 &  50 \\
0 vs 10 &  92.8 & 100 &  82.8 & 100 &  64.6 &  75 &  88.1 & 100 &  68.3 & 100 &  47.2 &  50 \\
0 vs 20 &  91.4 & 100 &  78.5 & 100 &  61.1 &  57 &  86.2 & 100 &  63.8 &  96 &  43.1 &  35 \\
5 vs 10 &  97.5 & 100 &  93.3 & 100 &  71.5 & 100 &  93.1 & 100 &  81.5 & 100 &  54.4 &  57 \\
5 vs 20 &  96.0 & 100 &  88.3 & 100 &  65.2 &  75 &  90.4 & 100 &  74.5 & 100 &  46.7 &  50 \\
10 vs 20 &  97.4 & 100 &  92.3 & 100 &  68.3 & 100 &  92.3 & 100 &  79.8 & 100 &  49.9 &  50 \\

\bottomrule
\end{tabular}
}
\vspace{1mm}
\caption{Storage slot overlap across state lookback distances $N$.
Read overlap denotes the Jaccard similarity of the read-sets at the two states, and write overlap is defined analogously for the write-sets. Value overlap denotes the fraction of overlapping written slots whose values are identical. Each metric is averaged over the transactions for which it is defined. The write and value populations are subsets of the read population.}
\label{tab:overlap}
\end{table}

Next, we analyze the mean and median overlap of storage slot accesses across execution states for Ethereum and Base. We consider overlap separately for storage slots that are read, slots that are written, and, conditional on overlapping writes, whether the values written are identical. Table~\ref{tab:overlap} reports these overlaps for different pairs of lookback distances. We note that the (landed, $N=0$) comparison is particularly significant: it represents the minimum realistic gap between simulation and execution, capturing only the state changes introduced by transactions preceding the given one within the same block. Even under this best-case scenario, the divergence on Base is clearly visible, and it is concentrated in the \emph{write} sets rather than the read sets.

In line with our previous analysis, transactions on Base exhibit greater state sensitivity than those on Ethereum. We compare the landed state to the start-of-block lookback state ($N=0$), which captures the effect of state modifications by preceding transactions within the same block. Read sets remain largely stable on both chains. The mean read overlap is 97.1\% on Ethereum and 93.5\% on Base, with a median of 100\% on both, meaning the typical transaction reads exactly the same slots. Write sets diverge far more. The mean write overlap is 94.6\% on Ethereum but only 80.8\% on Base, and the gap widens at larger lookback distances. At $N=20$, mean write overlap falls to 75.6\% on Ethereum and 59.3\% on Base.

Most notably, even when transactions write to the same storage slots, the values written do not necessarily coincide. On Ethereum at the (landed, $N=0$) pair, transactions mostly write to the same slots (mean write overlap 94.6\%, median 100\%). However, the written values often differ: mean value overlap is only 78.5\% (median 100\%), indicating that a non-trivial tail of transactions updates the same slot with different values. On Base this tail effect is stronger: mean value overlap of 67.8\%, with a median of 80\%. State sensitivity thus manifests at two distinct levels. Value divergence, i.e., writing different values to the same slots, reflects the expected dependency of computation on mutable state such as pool balances or oracle prices. Slot divergence, i.e., writing to a different set of storage locations, is more consequential, as it indicates that the transaction's execution path itself has changed under the altered pre-state. On Base, where write overlap at lookback 0 is already only 80.8\%, both forms of divergence are prevalent even within a single block.

One class of transactions for which divergent execution outcomes are expected are DEXes within DeFi. Because execution outcomes depend on mutable on-chain state, such as pool balances and prices, even small differences in the execution state can lead to different results. Consistent with this, even at lookback 0, MEV transactions on Base already exhibit write overlap of only 45.3\% and value overlap of 57.6\%, compared to 81.0\% and 65.0\% on Ethereum. At lookback 5, these drop further to 30.8\% and 38.0\% on Base versus 59.2\% and 35.9\% on Ethereum. DeFi transactions follow a similar pattern, with value overlap falling from 63.8\% at lookback 0 to 43.3\% at lookback 5 on Base, compared to 68.1\% and 51.8\% on Ethereum. The full category-level breakdown is reported in Appendix~\ref{app:slot-overlap-category}.

State drift grows monotonically with lookback distance. On Base, mean value overlap falls from 67.8\% at lookback 0 to 43.1\% at lookback 20; on Ethereum, from 78.5\% to 59.6\%. Two factors contribute: elapsed wall-clock time and the number of state-modifying transactions in the intervening blocks. Our data does not separately quantify these contributions, but the lookback-10 and lookback-20 measurements preview the simulation--execution gap that higher-throughput chains will increasingly encounter. In this regime, DeFi and MEV divergence will likely intensify.

Overall, while many transactions access largely stable sets of storage slots across lookbacks, a non-trivial subset exhibits substantial variability. As a result, both the storage accessed and the values written by a transaction are often difficult to predict in advance, complicating workload estimation and degrading transaction predictability from a user perspective.

%% file: suggestions.tex
\section{Concluding Discussion and Future Directions for Mitigation}
\label{sec:suggestions}

Our measurements reveal that EVM execution conditions are not stable: the resource mix varies across chains and shifts over time; persistent state growth is large and compositionally diverse; and execution outcomes (gas usage, storage access, success or failure) change when a transaction is re-evaluated on nearby historical states. Two assumptions underlying current gas metering are strained by these findings: that the workload mix is stable enough for one-dimensional gas pricing with fixed relative prices to remain appropriate, and that the state observed at simulation matches the state used at execution. Below, we first discuss how next-generation execution models sharpen both problems, then turn to mitigations.

\subsection{Future Designs Amplify Both Problems}

\paragraph*{Delayed and asynchronous execution.}
Under delayed or asynchronous execution~\cite{monadxyz2025,acp194}, state drift between submission and execution means gas-limit-based charging can systematically overbill users or cause reverts. Unlike synchronous execution, where overestimating the gas limit is largely harmless, asynchronous designs penalize over-declaration (e.g., charging on the full declared gas limit) because actual usage is not yet known at inclusion time~\cite{monadxyz2025}. State-sensitive transactions then either pay a margin to absorb the drift or risk failing.

\paragraph*{Optimistic parallel execution.}
Under optimistic parallel execution, diverging storage-access patterns across nearby states mean speculative caches are frequently invalidated on conflict-driven retry, potentially eroding parallelism's performance gains. Perhaps more importantly, statically declared access lists~\cite{eip2930,heimbach2023dissecting} face the same issue: if the slots a transaction touches depend on state, the list cannot be populated accurately ahead of time.

\paragraph*{Encrypted mempools and multiple concurrent proposers.}

Encrypted mempools widen the simulation-execution gap further: contending transactions remain hidden until decryption, so simulation cannot anticipate the writes that will land between submission and execution. Multiple concurrent proposers fragment block production across parties, so the ordering and pre-state of a transaction are not known until aggregation. Both widen the gap our measurements quantify between submission-time and execution-time state.

\subsection{Mitigations}

\paragraph*{Multi-dimensional fee markets.}

When the workload mix drifts over time, one-dimensional gas metering cannot adapt on its own. The only recourse is manual repricing. Multi-dimensional fee markets~\cite{diamandis2023designing,angeris2024multidimensional,eip8011} remove this rigidity by pricing distinct resources separately, letting prices for different resources converge to different equilibria. Our cross-chain and temporal measurements provide an empirical basis for the dimensions that matter: computation, storage reads, storage writes, call-related operations, and calldata.

\paragraph*{Explicit pricing of persistent state.}
Persistent state growth is large (435~GB on Base versus 30~GB on Ethereum in 2025), highly bursty (Table~\ref{tab:storage_distribution}), and compositionally diverse: storage slots dominate on both chains, bytecode accounts for roughly 25\% of Base's growth, and account births remain a material component, especially for simple transfers on Ethereum.
Because gas pricing treats state writes as transient per-block costs, it fails to reflect their permanent nature.
An explicit storage cost per byte (rather than gas), whether as ongoing rent or upfront deposits~\cite{buterin2024purge,buterin2021verkle}, should account for storage slots, contract bytecode, and account creation.

\paragraph*{Reducing state sensitivity.}
Gas estimates change across nearby historical states for 46\% of transactions on Base and 13.9\% on Ethereum, with DeFi and MEV exhibiting especially high sensitivity. Some of this sensitivity is intrinsic to state-dependent opcode costs (cold versus warm access, zero versus non-zero \texttt{SSTORE} transitions).
Given that sensitivity is structural, mitigations should target how the residual gap is billed rather than try to eliminate it. Under asynchronous execution, penalizing over-declaration is necessary for DoS protection, since actual usage is not yet known at inclusion time. Refunding a portion of the gap between declared limit and observed execution gas caps the overbilling penalty when a transaction lands on a more favorable state, while still discouraging inflated limits that reserve capacity the user does not need. For optimistic parallel execution, the category-level heterogeneity we observe suggests that schedulers can prioritize pre-warming and serialization for high-sensitivity categories (MEV, DeFi) while speculating freely on transfers and ERC-20 traffic, rather than a uniform speculation rule.

%% file: appendix.tex
\newpage
\section{Daily State Growth Distribution}\label{app:storage_distribution}

Table~\ref{tab:storage_distribution} reports the distribution of daily storage growth in 2025, broken down into storage slots, contract bytecode, and account state. For both chains, daily storage growth varies substantially, highlighting the mismatch between short-term per-block gas limits and the long-term cost of persistent state accumulation.

\begin{table*}[ht]
\centering
\resizebox{\textwidth}{!}{
\begin{tabular}{llrrrrrrrrr}
\toprule
\textbf{Chain} & \textbf{Metric} & \textbf{Mean} & \textbf{Std Dev} & \textbf{P1} & \textbf{P10} & \textbf{P25} & \textbf{P50} & \textbf{P75} & \textbf{P90} & \textbf{P99} \\
\midrule
Ethereum & Slots & 41.8 & 23.0 & 16.0 & 21.9 & 25.4 & 34.1 & 50.5 & 71.5 & 115.9 \\
 & Bytecode & 14.9 & 5.5 & 5.7 & 9.4 & 11.6 & 14.2 & 17.0 & 20.9 & 32.0 \\
 & Account State & 25.3 & 10.4 & 13.1 & 16.0 & 18.8 & 22.9 & 28.0 & 36.6 & 64.2 \\
 & Total & 82.0 & 36.4 & 37.9 & 49.8 & 57.4 & 71.4 & 96.4 & 127.5 & 209.0 \\
\midrule
Base & Slots & 732.6 & 309.3 & 176.5 & 348.8 & 567.1 & 709.2 & 852.1 & 1085.6 & 1767.0 \\
 & Bytecode & 303.0 & 247.9 & 74.0 & 95.5 & 121.5 & 191.6 & 419.7 & 644.8 & 1063.0 \\
 & Account State & 156.7 & 148.1 & 26.1 & 36.5 & 51.8 & 91.4 & 218.6 & 359.3 & 662.4 \\
 & Total & 1192.4 & 603.2 & 354.5 & 586.1 & 816.1 & 1030.9 & 1456.3 & 1999.0 & 3075.3 \\
\bottomrule
\end{tabular}}
\caption{Daily state storage growth distribution on Ethereum and Base in 2025. Values show storage slots (64 bytes each), contract bytecode, account state (128 bytes per net account transition), and total combined storage. All values are in MB/day.}\label{tab:storage_distribution}
\end{table*}

\section{Distribution of Activity Across Addresses}\label{app:address}

\begin{figure}[htb]
    \centering
    \begin{subfigure}{0.49\linewidth}
    \includegraphics[width=\linewidth]{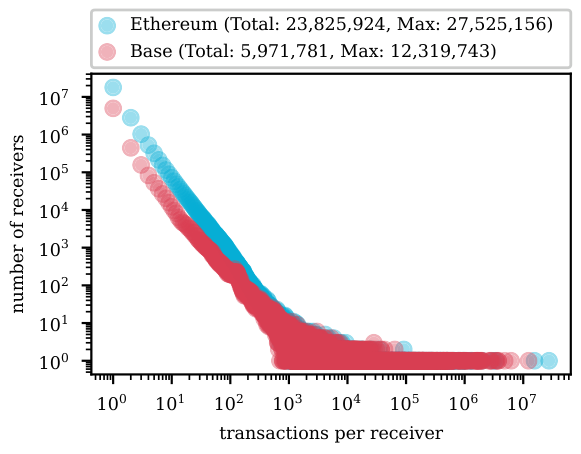}
    \caption{Receivers}\label{fig:powerlaw_comparison_receivers_txcount_overlay}
    \end{subfigure}\hfill
    \begin{subfigure}{0.49\linewidth}
    \includegraphics[width=\linewidth]{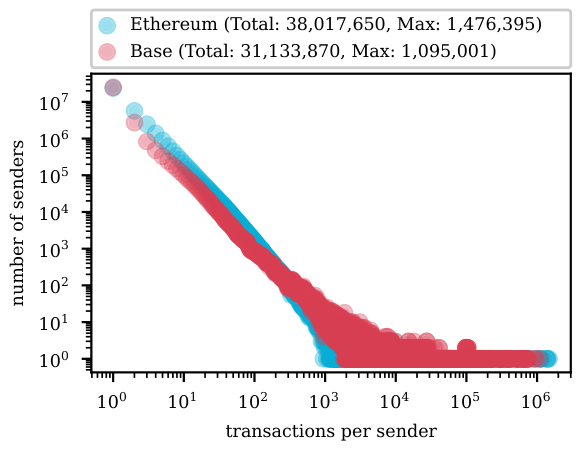}
    \caption{Senders}\label{fig:powerlaw_comparison_senders_txcount_overlay}
    \end{subfigure}

    \caption{Distribution of transaction activity per address on Ethereum and Base.
    Figure~\ref{fig:powerlaw_comparison_receivers_txcount_overlay} shows the number of transactions received per address, and Figure~\ref{fig:powerlaw_comparison_senders_txcount_overlay} shows the number of transactions sent per address.
    Contract creations (no receiver) and self-transfers are excluded.
    }\label{fig:powerlaw}
\end{figure}

At the address level, Figure~\ref{fig:powerlaw} shows the distribution of transaction activity across sender and receiver addresses on both chains. Both Ethereum and Base exhibit heavy-tailed distributions, in which most addresses participate in relatively few transactions while a small fraction of highly active addresses accounts for a large share of transaction volume. Within this population, which excludes contract creations and self-transfers, Ethereum shows higher address participation, with 23.8 million unique receivers and 38.0 million unique senders, compared to 6.0 million receivers and 31.1 million senders on Base. At the same time, activity on Base is more concentrated, with a larger share of transactions attributed to a smaller set of highly active addresses.

\section{Gas Decomposition by Category}\label{app:gas_by_category}

\begin{table*}[!tp]
\centering
\begin{subtable}{0.501\textwidth}
\centering
\resizebox{\textwidth}{!}{
\begin{tabular}{lrr}
\toprule
\textbf{Category} & \textbf{Ethereum} & \textbf{Base} \\
\midrule
\multicolumn{3}{l}{\textbf{MEV:}} \\
\quad Transactions & 14,458,276 & 111,896,672 \\
\quad Total Gas & 295,627 & 220,110 \\
\quad Execution & 265,182 & 196,080 \\
\quad Intrinsic & 52,282 & 25,768 \\
\quad Access List & 26,293 & 134 \\
\quad Authorization List & 3 & 0 \\
\quad Refund & 21,837 & 1,740 \\
\midrule
\multicolumn{3}{l}{\textbf{DeFi:}} \\
\quad Transactions & 31,854,577 & 33,704,931 \\
\quad Total Gas & 180,632 & 228,276 \\
\quad Execution & 172,888 & 226,063 \\
\quad Intrinsic & 29,102 & 26,924 \\
\quad Access List & 2,251 & 713 \\
\quad Authorization List & 67 & 62 \\
\quad Refund & 21,365 & 24,712 \\
\midrule
\multicolumn{3}{l}{\textbf{Token:}} \\
\quad Transactions & 63,950,730 & 20,354,249 \\
\quad Total Gas & 63,563 & 174,677 \\
\quad Execution & 43,961 & 157,142 \\
\quad Intrinsic & 21,805 & 22,888 \\
\quad Access List & 8 & 0 \\
\quad Authorization List & 0 & 0 \\
\quad Refund & 2,203 & 5,353 \\
\midrule
\multicolumn{3}{l}{\textbf{Contract Creation:}} \\
\quad Transactions & 215,137 & 214,308 \\
\quad Total Gas & 1,670,843 & 1,128,603 \\
\quad Execution & 710,574 & 196,465 \\
\quad Intrinsic & 150,530 & 130,242 \\
\quad Access List & 1,181 & 0 \\
\quad Authorization List & 0 & 0 \\
\quad Refund & 2,024 & 797 \\
\midrule
\multicolumn{3}{l}{\textbf{Infrastructure:}} \\
\quad Transactions & 3,382,961 & 7,368,706 \\
\quad Total Gas & 210,106 & 792,576 \\
\quad Execution & 201,168 & 790,018 \\
\quad Intrinsic & 32,785 & 35,860 \\
\quad Access List & 77 & 0 \\
\quad Authorization List & 1,484 & 24 \\
\quad Refund & 23,847 & 33,302 \\
\bottomrule
\end{tabular}
}
\caption{Per-Transaction Gas Statistics}\label{tab:cat_per_tx}
\end{subtable}
\hfill
\begin{subtable}{0.4\textwidth}
\centering
\resizebox{\textwidth}{!}{
\begin{tabular}{lrr}
\toprule
\textbf{Category} & \textbf{Ethereum} & \textbf{Base} \\
\midrule
\multicolumn{3}{l}{\textbf{MEV:}} \\
\quad Storage Read & 18.1 & 36.5 \\
\quad Storage Write & 37.0 & 8.3 \\
\quad Compute & 23.9 & 30.9 \\
\quad Calls & 7.8 & 21.0 \\
\quad Contract Ops & 6.6 & 2.1 \\
\quad Logging & 6.3 & 1.0 \\
\quad Other & 0.3 & 0.1 \\
\midrule
\multicolumn{3}{l}{\textbf{DeFi:}} \\
\quad Storage Read & 25.1 & 26.3 \\
\quad Storage Write & 31.5 & 30.7 \\
\quad Compute & 22.0 & 23.8 \\
\quad Calls & 11.2 & 8.7 \\
\quad Contract Ops & 3.0 & 4.0 \\
\quad Logging & 6.9 & 6.3 \\
\quad Other & 0.2 & 0.2 \\
\midrule
\multicolumn{3}{l}{\textbf{Token:}} \\
\quad Storage Read & 30.6 & 17.6 \\
\quad Storage Write & 45.7 & 50.9 \\
\quad Compute & 9.2 & 11.3 \\
\quad Calls & 3.2 & 3.2 \\
\quad Contract Ops & 5.2 & 13.4 \\
\quad Logging & 5.4 & 3.6 \\
\quad Other & 0.7 & 0.0 \\
\midrule
\multicolumn{3}{l}{\textbf{Contract Creation:}} \\
\quad Storage Read & 9.7 & 5.6 \\
\quad Storage Write & 53.7 & 50.9 \\
\quad Compute & 9.5 & 4.1 \\
\quad Calls & 0.8 & 0.5 \\
\quad Contract Ops & 6.5 & 23.1 \\
\quad Logging & 18.7 & 4.1 \\
\quad Other & 1.0 & 11.7 \\
\midrule
\multicolumn{3}{l}{\textbf{Infrastructure:}} \\
\quad Storage Read & 19.8 & 20.0 \\
\quad Storage Write & 31.9 & 41.9 \\
\quad Compute & 23.9 & 20.1 \\
\quad Calls & 16.3 & 6.0 \\
\quad Contract Ops & 2.2 & 6.9 \\
\quad Logging & 5.5 & 4.9 \\
\quad Other & 0.4 & 0.2 \\
\bottomrule
\end{tabular}
}
\caption{Execution Gas Breakdown (\%)}\label{tab:cat_exec_gas}
\end{subtable}
\caption{Gas usage by transaction category on Ethereum and Base in 2025.
Table~\ref{tab:cat_per_tx} reports per-transaction gas statistics for five categories selected to span distinct workload shapes, including intrinsic, execution, access list, authorization list, and refund components.
Table~\ref{tab:cat_exec_gas} reports the execution gas breakdown by EVM operation category for each address class.}
\label{tab:gas_by_category}
\end{table*}

Table~\ref{tab:gas_by_category} shows the decomposition of gas usage by transaction category for Ethereum and Base in 2025. For each category, Table~\ref{tab:cat_per_tx} summarizes per-transaction gas statistics, including intrinsic gas, execution gas, access list and authorization list gas, as well as gas refunds.

Table~\ref{tab:cat_exec_gas} further decomposes execution gas by opcode family: storage reads and writes, computation, call overhead, contract-related operations, logging, and residual operations.

Across categories, execution gas constitutes the dominant share of per-transaction gas usage on both chains. MEV transactions are read-heavy on Base (36.5\% storage reads, 8.3\% storage writes) but write-heavy on Ethereum (18.1\% storage reads, 37.0\% storage writes). Infrastructure transactions on Base exhibit a strongly write-heavy profile, with storage writes accounting for 41.9\% of execution gas, compared to 31.9\% on Ethereum. Token transactions are write-intensive on both chains, with storage writes contributing 45.7\% of execution gas on Ethereum and 50.9\% on Base. Contract creation is dominated by storage writes, which account for 53.7\% of execution gas on Ethereum and 50.9\% on Base, reflecting the persistent state a deployment establishes. Contract-related operations form the second component and differ sharply across chains, at 23.1\% on Base against 6.5\% on Ethereum, while logging accounts for 18.7\% on Ethereum and only 4.1\% on Base.

\begin{table}[t!]
\centering

\begin{subtable}{\textwidth}
\centering
\resizebox{\textwidth}{!}{
\begin{tabular}{lrrrrrrrrrrrr}
\toprule
& & & \multicolumn{4}{c}{\textbf{CV (\%)}} & \multicolumn{5}{c}{\textbf{LB (blocks)}} \\
\cmidrule(lr){4-7} \cmidrule(lr){8-12}
\textbf{Category} & \textbf{Txs} & \textbf{NZ (\%)} & \textbf{Mean} & \textbf{P50} & \textbf{P90} & \textbf{P99} & \textbf{P1} & \textbf{P5} & \textbf{P10} & \textbf{P25} & \textbf{P50} \\
\midrule
DeFi & 2,669,074 & 44.0 & 1.20 & 0.00 & 3.32 & 13.20 & 0 & 1 & 2 & 9 & 20 \\
MEV & 654,443 & 61.2 & 1.81 & 0.89 & 4.13 & 13.85 & 0 & 0 & 0 & 7 & 20 \\
Token & 5,965,370 & 6.2 & 0.36 & 0.00 & 0.00 & 15.23 & 0 & 1 & 3 & 15 & 20 \\
Infrastructure & 268,566 & 41.2 & 3.81 & 0.00 & 14.24 & 61.54 & 0 & 1 & 3 & 18 & 20 \\
Simple Transfer & 6,710,164 & 3.0 & 0.01 & 0.00 & 0.00 & 0.40 & 0 & 1 & 2 & 15 & 20 \\
Contract Creation & 18,050 & 0.8 & 0.01 & 0.00 & 0.00 & 0.00 & 0 & 0 & 1 & 17 & 20 \\
Bridge & 474,549 & 16.7 & 0.32 & 0.00 & 0.21 & 6.00 & 0 & 0 & 1 & 13 & 20 \\
NFT & 232,040 & 8.6 & 0.40 & 0.00 & 0.00 & 10.20 & 0 & 1 & 2 & 10 & 20 \\
CEX & 154,236 & 14.6 & 1.04 & 0.00 & 1.37 & 18.43 & 1 & 2 & 4 & 9 & 20 \\
L2 & 128,350 & 0.1 & 0.00 & 0.00 & 0.00 & 0.00 & 3 & 19 & 20 & 20 & 20 \\
Other & 1,987,417 & 14.6 & 1.51 & 0.00 & 0.79 & 33.07 & 0 & 1 & 4 & 20 & 20 \\
\bottomrule
\end{tabular}
}
\caption{Ethereum}
\label{tab:estimate_variance_by_category_ethereum}
\end{subtable}

\begin{subtable}{\textwidth}
\centering
\resizebox{\textwidth}{!}{
\begin{tabular}{lrrrrrrrrrrrr}
\toprule
& & & \multicolumn{4}{c}{\textbf{CV (\%)}} & \multicolumn{5}{c}{\textbf{LB (blocks)}} \\
\cmidrule(lr){4-7} \cmidrule(lr){8-12}
\textbf{Category} & \textbf{Txs} & \textbf{NZ (\%)} & \textbf{Mean} & \textbf{P50} & \textbf{P90} & \textbf{P99} & \textbf{P1} & \textbf{P5} & \textbf{P10} & \textbf{P25} & \textbf{P50} \\
\midrule
DeFi & 2,580,943 & 53.0 & 4.02 & 0.15 & 4.82 & 86.07 & 0 & 1 & 3 & 17 & 20 \\
MEV & 11,273,037 & 54.2 & 10.11 & 0.01 & 33.07 & 155.65 & 5 & 20 & 20 & 20 & 20 \\
Token & 1,207,871 & 9.4 & 0.77 & 0.00 & 0.00 & 17.87 & 0 & 2 & 4 & 20 & 20 \\
Infrastructure & 550,241 & 12.0 & 2.31 & 0.00 & 0.07 & 83.60 & 1 & 5 & 19 & 20 & 20 \\
Simple Transfer & 1,459,442 & 0.0 & 0.00 & 0.00 & 0.00 & 0.00 & 1 & 3 & 5 & 15 & 20 \\
Contract Creation & 31,638 & 0.0 & 0.00 & 0.00 & 0.00 & 0.00 & 2 & 10 & 20 & 20 & 20 \\
Bridge & 128,224 & 19.8 & 2.70 & 0.00 & 8.67 & 34.12 & 0 & 1 & 1 & 20 & 20 \\
NFT & 290,174 & 19.7 & 1.45 & 0.00 & 6.33 & 18.43 & 1 & 2 & 4 & 8 & 20 \\
CEX & 6,471 & 7.4 & 0.02 & 0.00 & 0.00 & 0.18 & 5 & 20 & 20 & 20 & 20 \\
Other & 4,892,937 & 52.4 & 5.41 & 0.01 & 8.26 & 149.24 & 0 & 2 & 7 & 20 & 20 \\
\bottomrule
\end{tabular}
}
\caption{Base}
\label{tab:estimate_variance_by_category_base}
\end{subtable}
\caption{Gas estimate variance by transaction category for (\ref{tab:estimate_variance_by_category_ethereum}) Ethereum and (\ref{tab:estimate_variance_by_category_base}) Base in September 2025.
CV denotes the coefficient of variation (standard deviation divided by mean, expressed as a percentage) of gas estimates across state lookback distances. CV is not bounded by 100\%. Values above 100\% indicate that the standard deviation of a transaction's gas estimates exceeds their mean.
Max Successful Simulation Lookback reports the maximum lookback distance (in blocks), up to 20, at which transaction simulation succeeds without error.}
\label{tab:estimate_variance_by_category}
\end{table}
\section{Gas Estimate Variance by Transaction Category}\label{app:estimate_variance_category}

Table~\ref{tab:estimate_variance_by_category} reports gas estimate variance by transaction category. DeFi and MEV transactions exhibit the highest fraction of non-zero variance on both chains, while simple transfers show near-zero variance on Base but non-trivial variance on Ethereum due to EIP-7702.

\section{Gas Variance by Transaction Category and Opcode Group}\label{app:gas_variance_category}

Table~\ref{tab:gas_variance_category_opcode} reports, by transaction category and opcode group, the mean coefficient of variation of each opcode group's share of total transaction gas across lookbacks. MEV transactions on Base exhibit the highest variability of any Base category, reaching mean share CVs of 11.45\% for contract-related operations, 9.40\% for storage writes, and 5.55\% for logging, with \textit{Infrastructure} close behind (9.11\% and 8.21\% respectively). DeFi is more moderate (up to 4.33\%), and \textit{Simple Transfer}, \textit{Contract Creation}, and \textit{CEX} are effectively invariant on Base (all below 0.06\%). On Ethereum, variability is lower for most categories, but \textit{Infrastructure} is a marked exception at a 15.91\% mean share CV for storage writes, the largest value in the table, followed by \textit{CEX} logging at 8.72\%.

\begin{table}[ht]
\centering
\resizebox{\textwidth}{!}{
\begin{tabular}{lrrrrrrrrrrrr}
\toprule
& \multicolumn{6}{c}{\textbf{Ethereum}} & \multicolumn{6}{c}{\textbf{Base}} \\
\cmidrule(lr){2-7} \cmidrule(lr){8-13}
\textbf{Category} & \textbf{S-Read} & \textbf{S-Write} & \textbf{Compute} & \textbf{Calls} & \textbf{Contract} & \textbf{Log} & \textbf{S-Read} & \textbf{S-Write} & \textbf{Compute} & \textbf{Calls} & \textbf{Contract} & \textbf{Log} \\
\midrule
DeFi & 0.89 & 1.71 & 0.93 & 1.14 & 1.12 & 1.29 & 2.04 & 2.92 & 2.09 & 4.33 & 4.14 & 2.68 \\
MEV & 1.92 & 3.01 & 1.25 & 2.38 & 2.19 & 2.62 & 3.43 & 9.40 & 3.87 & 4.30 & 11.45 & 5.55 \\
Token & 0.37 & 1.10 & 0.44 & 0.21 & 0.06 & 0.46 & 0.35 & 1.05 & 0.52 & 0.40 & 0.23 & 0.50 \\
Infrastructure & 3.13 & 15.91 & 2.15 & 5.28 & 4.45 & 5.30 & 2.99 & 8.21 & 2.46 & 4.40 & 9.11 & 2.57 \\
Simple Transfer & 0.00 & 0.00 & 0.00 & 0.00 & 0.00 & 0.00 & 0.00 & 0.00 & 0.00 & 0.00 & 0.00 & 0.00 \\
Contract Creation & 0.04 & 0.02 & 0.01 & 0.05 & 0.00 & 0.03 & 0.00 & 0.00 & 0.00 & 0.00 & 0.01 & 0.00 \\
Bridge & 0.30 & 0.55 & 0.28 & 0.41 & 0.31 & 0.25 & 1.42 & 8.38 & 1.03 & 3.89 & 2.51 & 2.88 \\
NFT & 0.45 & 0.81 & 0.26 & 0.36 & 0.48 & 0.58 & 3.50 & 1.87 & 0.72 & 2.63 & 1.66 & 1.87 \\
CEX & 0.83 & 8.41 & 1.62 & 1.54 & 0.79 & 8.72 & 0.03 & 0.05 & 0.03 & 0.03 & 0.03 & 0.03 \\
L2 & 0.01 & 0.00 & 0.01 & 0.00 & 0.00 & 0.00 & -- & -- & -- & -- & -- & -- \\
Other & 1.93 & 5.52 & 2.65 & 2.06 & 1.79 & 4.69 & 3.07 & 5.59 & 3.95 & 3.73 & 4.71 & 5.00 \\
\bottomrule
\end{tabular}
}
\vspace{1mm}
\caption{Gas variance by transaction category and opcode group. Shows the mean coefficient of variation (CV\%) of each opcode group's share of total transaction gas across lookbacks.}
\label{tab:gas_variance_category_opcode}
\end{table}

\section{Storage Slots Accessed per Transaction Across Lookbacks}
\label{app:slots-per-tx}

\begin{table}[ht!]
\centering

\resizebox{\textwidth}{!}{
\begin{tabular}{lrrrrrrrrrrrrrrrr}
\toprule
& \multicolumn{8}{c}{\textbf{Ethereum}} & \multicolumn{8}{c}{\textbf{Base}} \\
\cmidrule(lr){2-9} \cmidrule(lr){10-17}
\textbf{LB} & \textbf{Mean} & \textbf{Std} & \textbf{P10} & \textbf{P25} & \textbf{P50} & \textbf{P75} & \textbf{P90} & \textbf{P99} & \textbf{Mean} & \textbf{Std} & \textbf{P10} & \textbf{P25} & \textbf{P50} & \textbf{P75} & \textbf{P90} & \textbf{P99} \\
\midrule
 Landed & 14.66 & 34.05 &    2 &    5 &    7 &   17 &   29 &  103 & 22.02 & 65.24 &    2 &    4 &    8 &   21 &   52 &  178 \\
  0 & 13.44 & 30.43 &    2 &    5 &    7 &   15 &   28 &   88 & 21.89 & 64.72 &    2 &    4 &    9 &   21 &   50 &  178 \\
  5 & 12.39 & 29.86 &    2 &    4 &    7 &   14 &   25 &   81 & 20.36 & 63.06 &    2 &    4 &    8 &   19 &   45 &  166 \\
 10 & 12.11 & 29.56 &    2 &    4 &    7 &   13 &   25 &   80 & 19.78 & 60.69 &    2 &    4 &    8 &   19 &   44 &  160 \\
 20 & 11.79 & 29.29 &    2 &    4 &    7 &   13 &   24 &   78 & 19.21 & 58.28 &    2 &    4 &    7 &   18 &   42 &  152 \\

\bottomrule
\end{tabular}
}
\vspace{1mm}
\caption{Distribution of the total number of unique storage slots accessed per transaction.
Transactions are re-executed on historical states at lookback (LB) distances $N$, where execution uses the state of block $b - 1 - N$ for a transaction included in block $b$.
The landed row is the state at which the transaction actually executed. Total slots includes both the readset and the writeset.}
\label{tab:per_tx_slots}
\end{table}

Table~\ref{tab:per_tx_slots} reports the distribution of the number of accessed storage slots across lookback distances. The analysis includes only transactions that access at least one storage slot; transactions with no storage accesses are excluded.

At the landed state, transactions on Ethereum access an average of 14.66 storage slots, compared to 22.02 on Base. The standard deviation is likewise higher on Base (65.24) than on Ethereum (34.05), indicating substantially greater variability in accessed slots. At the 99th percentile, transactions access 103 slots on Ethereum and 178 slots on Base.

Across both chains, the number of accessed storage slots decreases as the lookback distance increases. This pattern likely reflects selection effects, as transactions that remain executable at larger lookback distances tend to be less state-sensitive and thus access fewer storage slots.

\section{Slot Overlap by Transaction Category}
\label{app:slot-overlap-category}

Table~\ref{tab:per_category_overlap} reports storage slot overlap by transaction category, comparing the landed state to historical lookback states at various distances. On Ethereum, DeFi transactions already exhibit relatively low value overlap. For example, for the (landed, $N=5$) pair, the mean value overlap for DeFi is 51.8\% on Ethereum and 43.3\% on Base. MEV transactions diverge even more strongly: on Ethereum, their write overlap (59.2\% at (landed, $N=5$)) and value overlap (35.9\%) are both markedly below the DeFi values, indicating that MEV transactions both write to different state locations and write different values as the execution state shifts.

On Base, this effect is far more pronounced. MEV transactions exhibit the lowest write overlap of any major category, at only 45.3\% even for the (landed, $N=0$) pair and 30.8\% at (landed, $N=5$), roughly half the corresponding Ethereum values. This suggests that, as analyzed in previous work, MEV execution on Base is more heavily optimistic, with execution outcomes depending more strongly on the precise execution state~\cite{babel2023clockwork}.

\begin{table}[ht!]
\centering
\begin{subtable}{\textwidth}
\centering
\resizebox{\textwidth}{!}{
\begin{tabular}{lrrrrrrrrrrrrrrrrrrrrrr}
\toprule
Category & Txs & \multicolumn{3}{c}{Landed vs 0} & \multicolumn{3}{c}{Landed vs 5} & \multicolumn{3}{c}{Landed vs 10} & \multicolumn{3}{c}{Landed vs 20} & \multicolumn{3}{c}{0 vs 5} & \multicolumn{3}{c}{0 vs 10} & \multicolumn{3}{c}{0 vs 20} \\
 \cmidrule(lr){3-5} \cmidrule(lr){6-8} \cmidrule(lr){9-11} \cmidrule(lr){12-14} \cmidrule(lr){15-17} \cmidrule(lr){18-20} \cmidrule(lr){21-23}
 & & Read & Write & Value & Read & Write & Value & Read & Write & Value & Read & Write & Value & Read & Write & Value & Read & Write & Value & Read & Write & Value \\
\midrule
DeFi & 2,853,038 &  95.0 &  89.9 &  68.1 &  85.7 &  73.3 &  51.8 &  83.2 &  68.6 &  48.1 &  80.9 &  64.1 &  45.4 &  88.9 &  78.1 &  53.9 &  86.5 &  73.0 &  49.3 &  84.1 &  68.2 &  46.3 \\
MEV & 749,266 &  88.9 &  81.0 &  65.0 &  80.5 &  59.2 &  35.9 &  78.4 &  55.6 &  29.3 &  76.6 &  52.2 &  23.9 &  88.8 &  68.6 &  37.5 &  86.7 &  64.5 &  30.2 &  84.9 &  60.6 &  24.4 \\
Token & 5,952,881 &  99.8 &  99.4 &  83.7 &  98.9 &  94.8 &  73.3 &  98.4 &  92.0 &  71.0 &  97.8 &  87.9 &  68.2 &  99.0 &  95.1 &  78.3 &  98.5 &  92.4 &  74.4 &  97.8 &  88.2 &  70.5 \\
Infrastructure & 283,595 &  96.0 &  93.2 &  76.9 &  81.4 &  69.0 &  67.3 &  78.8 &  64.8 &  65.8 &  76.0 &  60.1 &  64.7 &  84.9 &  72.8 &  69.2 &  82.1 &  68.2 &  67.3 &  79.4 &  63.3 &  66.1 \\
Contract Creation & 13,746 &  97.2 &  97.2 &  99.0 &  54.0 &  52.7 &  98.0 &  49.3 &  47.6 &  97.9 &  44.7 &  42.9 &  97.8 &  54.2 &  52.9 &  98.0 &  49.5 &  47.8 &  97.8 &  44.9 &  43.1 &  97.8 \\
Bridge & 524,055 &  92.1 &  87.3 &  86.8 &  84.0 &  70.3 &  67.0 &  82.7 &  67.6 &  62.7 &  81.0 &  63.7 &  60.2 &  91.7 &  81.1 &  66.9 &  90.5 &  78.4 &  61.7 &  88.9 &  74.5 &  58.5 \\
NFT & 189,829 &  92.5 &  85.5 &  81.7 &  86.9 &  73.8 &  75.0 &  87.1 &  72.8 &  73.1 &  86.5 &  71.3 &  71.7 &  88.1 &  75.0 &  75.4 &  87.8 &  73.6 &  73.4 &  87.0 &  71.9 &  72.0 \\
CEX & 106,310 &  99.2 &  98.1 &  95.1 &  97.3 &  91.5 &  86.0 &  96.0 &  85.5 &  82.0 &  95.7 &  83.7 &  76.6 &  98.1 &  93.0 &  87.7 &  96.7 &  87.0 &  83.2 &  96.4 &  85.1 &  77.4 \\
L2 & 33,346 &  64.5 &  45.2 &  79.8 &  62.8 &  40.7 &  78.7 &  61.7 &  38.9 &  78.8 &  56.4 &  26.2 &  85.1 &  97.8 &  89.9 &  78.8 &  95.2 &  85.9 &  78.8 &  86.9 &  57.4 &  85.2 \\
Other & 1,598,814 &  97.6 &  95.7 &  78.0 &  91.7 &  82.0 &  62.6 &  89.6 &  76.8 &  59.5 &  87.9 &  72.7 &  57.0 &  93.0 &  83.8 &  63.6 &  90.9 &  78.4 &  60.2 &  89.1 &  74.3 &  57.5 \\
\bottomrule
\end{tabular}
}
\caption{Ethereum}

\end{subtable}

\begin{subtable}{\textwidth}
\centering
\resizebox{\textwidth}{!}{
\begin{tabular}{lrrrrrrrrrrrrrrrrrrrrrr}
\toprule
Category & Txs & \multicolumn{3}{c}{Landed vs 0} & \multicolumn{3}{c}{Landed vs 5} & \multicolumn{3}{c}{Landed vs 10} & \multicolumn{3}{c}{Landed vs 20} & \multicolumn{3}{c}{0 vs 5} & \multicolumn{3}{c}{0 vs 10} & \multicolumn{3}{c}{0 vs 20} \\
 \cmidrule(lr){3-5} \cmidrule(lr){6-8} \cmidrule(lr){9-11} \cmidrule(lr){12-14} \cmidrule(lr){15-17} \cmidrule(lr){18-20} \cmidrule(lr){21-23}
 & & Read & Write & Value & Read & Write & Value & Read & Write & Value & Read & Write & Value & Read & Write & Value & Read & Write & Value & Read & Write & Value \\
\midrule
DeFi & 2,868,629 &  92.7 &  87.9 &  63.8 &  83.5 &  74.6 &  43.3 &  80.4 &  70.6 &  37.8 &  77.7 &  66.2 &  32.7 &  87.6 &  79.9 &  45.8 &  84.3 &  75.6 &  39.6 &  81.3 &  70.8 &  34.1 \\
MEV & 12,517,895 &  95.4 &  45.3 &  57.6 &  92.8 &  30.8 &  38.0 &  91.9 &  27.9 &  35.6 &  90.9 &  25.1 &  33.2 &  93.4 &  41.0 &  34.5 &  92.3 &  36.4 &  30.4 &  91.1 &  32.1 &  27.3 \\
Token & 1,376,774 &  95.4 &  91.2 &  83.3 &  91.6 &  84.6 &  73.5 &  90.8 &  83.3 &  71.6 &  90.1 &  81.6 &  70.0 &  95.1 &  90.4 &  77.6 &  94.0 &  88.5 &  75.1 &  92.9 &  86.2 &  73.2 \\
Infrastructure & 556,227 &  91.9 &  89.5 &  70.7 &  83.1 &  76.7 &  60.7 &  80.2 &  72.3 &  58.4 &  77.3 &  68.3 &  56.3 &  83.6 &  77.3 &  61.4 &  80.4 &  72.6 &  58.7 &  77.5 &  68.4 &  56.5 \\
Contract Creation & 20,199 &  98.2 &  97.8 &  99.7 &  68.8 &  70.7 &  99.6 &  53.2 &  56.4 &  99.5 &  44.4 &  46.6 &  99.3 &  69.5 &  71.6 &  99.6 &  53.9 &  57.3 &  99.5 &  45.1 &  47.6 &  99.3 \\
Bridge & 143,414 &  94.0 &  88.2 &  89.4 &  80.8 &  62.3 &  72.3 &  79.1 &  59.6 &  66.7 &  77.6 &  57.1 &  61.2 &  86.4 &  72.4 &  74.1 &  84.7 &  69.6 &  68.4 &  83.2 &  67.0 &  62.8 \\
NFT & 438,160 &  90.0 &  77.6 &  91.8 &  73.5 &  53.5 &  88.9 &  72.2 &  54.4 &  86.6 &  74.0 &  59.5 &  84.0 &  77.7 &  59.5 &  89.6 &  74.7 &  57.6 &  87.1 &  75.5 &  61.3 &  84.3 \\
CEX & 6,521 &  84.3 &  70.6 &  53.6 &  84.1 &  69.7 &  45.8 &  84.0 &  69.4 &  43.1 &  83.3 &  68.0 &  37.9 &  84.0 &  69.8 &  47.1 &  83.9 &  69.2 &  44.0 &  83.2 &  67.8 &  38.5 \\
Other & 5,428,723 &  89.6 &  82.0 &  64.3 &  79.6 &  66.0 &  44.2 &  77.0 &  62.2 &  40.6 &  74.5 &  58.0 &  37.0 &  83.7 &  73.5 &  45.9 &  80.2 &  68.7 &  41.0 &  77.1 &  63.7 &  36.9 \\
\bottomrule
\end{tabular}
}
\caption{Base}

\end{subtable}

\vspace{1mm}
\caption{Storage slot overlap by transaction category.
Results are shown for the most frequent categories on each chain. Smaller labeled categories and unlabeled transactions are folded into Other.
Read, write, and value columns report mean overlap percentages.
Transaction counts refer to the read-qualified population comparing the actual landed pre-state with the start-of-block state ($N=0$). Write and value averages cover only transactions where the metric is defined. As in Table~\ref{tab:per_tx_slots}, transactions that access no storage slot are excluded, which removes \textit{Simple Transfer} from this table entirely.}
\label{tab:per_category_overlap}
\end{table}